\documentclass{aa}
\usepackage{graphicx,epsfig}
\usepackage{amssymb,amsmath}
\usepackage{xspace}
\usepackage{natbib}
\usepackage{ulem}
\newcommand{\etal}{{et al.\ }}
\newcommand{\lta}{\stackrel{<}{\scriptstyle\sim}}
\newcommand{\gta}{\stackrel{>}{\scriptstyle\sim}}

\begin{document}

%
\title{Secondary Globular Cluster Populations} 

\author{Uta Fritze -- v. Alvensleben}
\institute{Universit\"ats-Sternwarte, Geismarlandstr.~11, 37083 G\"ottingen, Germany}
\authorrunning{Uta Fritze -- v. Alvensleben}
\titlerunning{Secondary Globular Cluster Populations}
\offprints{U. Fritze -- v. A., \email{ufritze@uni-sw.gwdg.de}}
\date{Received xxx  / Accepted xxx}

\abstract{
This study is motivated by two facts: 1. the formation of populous star cluster systems is widely observed to accompany violent star formation episodes in gas-rich galaxies as e.g. those triggered by strong interactions or merging. 2. The Globular Cluster ({\bf GC}) systems of most but not all early-type galaxies show bimodal optical color distributions with fairly universal blue peaks and somewhat variable red peak colors, yet their Luminosity Functions ({\bf LF}s) look like simple Gaussians with apparently universal turn-over magnitudes that are used for distance measurements and the determination of ${\rm H_o}$. 
Based on a new set of evolutionary synthesis models for Simple (= single burst) Stellar Populations ({\bf SSP}s) of various metallicities using the latest Padova isochrones I study the color and luminosity evolution of GC populations over the wavelength range from U through K, providing an extensive grid of models for comparison with observations. I assume the intrinsic widths of the color distributions and LFs to be constant in time at the values observed today for the Milky Way or M31 halo GC populations. Taking the color distributions and LFs of the Milky Way or M31 halo GC population as a reference for old metal-poor GC populations in general, I study for which combinations of age and metallicity a secondary GC population formed in some violent star formation event in the history of its parent galaxy  may or may not be detected in the observed GC color distributions. I also investigate the effect of these secondary GCs on the LFs of the total GC system. Significant differences are found among the diagnostic efficiencies in various wavelength regions. In particular, we predict the NIR to be able to reveal the presence of GC subpopulations with different age -- metallicity combinations that may perfectly hide within one inconspicuous optical color peak. 
If the entire manifold of possible age -- metallicity combinations is admitted for a secondary GC population, we find several cases where the resulting LF of the whole GC system is significantly affected and its turn-over could not serve as a reliable distance indicator. If, on the other hand, we assume some age -- metallicity relation for GC populations, the second peak of the LFs vanishes and models indicate single-peak GC LFs even in GC systems with bimodal color distributions. A broad but sufficient age -- metallicity relation is, for example, obtained if the secondary GC populations form in mergers of various spiral galaxy types from the ISM pre-enriched over the redshift range from z $\gta 4.4$ to z $\gta 0$. 
As a first illustrative example we apply our models to V- and I-band data presented by Larsen \etal (2001) for blue and red peak GCs in three early-type galaxies. We point out the importance of having multi-band information to independently constrain ages and metallicities of different GC subpopulations and again stress the diagnostic potential of K-band data in addition to optical observations. 
The models presented here will be used for the interpretation of multi-wavelength data on GC systems in galaxies of various types, luminosities and environments as well as on young star cluster systems in interacting galaxies and mergers. By independently constraining ages and metallicities of individual clusters we expect to shed light on both cluster and galaxy formation scenarios. 

}

\maketitle
%

\section{Introduction}
In the local Universe, we can witness the formation of -- sometimes populous -- star cluster systems in the powerful starbursts accompanying many gas-rich spiral galaxy mergers (cf. Schweizer 2002 for a recent review). Many of these clusters seem compact (cf. Carlson \& Holtzman, 2001) and massive enough (cf. Fritze -- v. Alvensleben 1999a) to be Globular Clusters ({\bf GC}s) able to survive for many Gyr. Formed from pre-enriched gas they will keep standing out in metallicity even at times when their colors and luminosities will have come close to those of old GCs. Spirals were more gas-rich in the past and mergers were more frequent (e.g. Le F\`evre \etal 2000). Hence, the formation of populous secondary GC systems may be expected to have happened almost all over the age of the Universe. 

The origin of elliptical galaxies is still a matter of debate. Were their stars formed ``all at once'' in the early Universe -- either in a monolithic collapse scenario or in 
some ``coordinated effort'' in building blocks that later merged together in a largely stellar-dynamical process -- or did they have significant populations of stars formed in one major spiral-spiral merger or in a series of hierarchical mergers that still involved enough gas to power star formation? 

Observations of metallicity and color distributions of GC systems may give us direct observational evidence about the formation history of their parent galaxies, even more so than the galaxies' integrated light (cf. S. Zepf 2002). Spectroscopy of reasonable numbers of GCs in elliptical or S0 galaxies out to Virgo cluster distances is underway with 10m class telescopes, but it has to cope with crowding and the bright and spatially variable galaxy backgrounds. GC (${\rm V-I}$) color distributions from HST observations are becoming available for large numbers of galaxies (e.g. Gebhardt \& Kissler -- Patig 1999, Larsen \etal 2001). The same is true for ground-based GC color distributions in the Washington system, which is more sensitive in disentangling ages and metallicities of old GCs (e.g. Geisler \etal 1996, Ostrov \etal 1998). Although being significantly less time consuming, broad band and even Washington color distributions of GC systems give less direct information about their age and metallicity distributions than spectroscopy. The reason is the well-known age -- metallicity degeneracy of colors (e.g. Worthey 1994). Empirical calibrations of metallicity in terms of colors are usually obtained for old GCs in the Milky Way ({\bf MW}) or M31 halos and strictly valid only over their metallicity range ${\rm -2.3 \leq [Fe/H] \leq -0.5}$ (e.g. Couture \etal 1990, Barmby \etal 2000). 

For most luminous elliptical and S0 galaxies, the observed GC color distributions are bimodal (Gebhard \& Kissler -- Patig 1999, Larsen \etal 2001, Kundu \& Whitmore 2001a, b), similar to the color distributions of the halo field stars in the case of NGC 3115 (Elson 1997, Kundu \& Witmore 1998) and NGC 5128 (Harris \etal 2000). Only a few low-luminosity E/S0 galaxies, like NGC 4478, show single-peak color distributions (Gebhardt \& Kissler -- Patig 1999), central cluster galaxies sometimes feature broad or multi-peak distributions (Forbes \etal 1997). Gaussians fitted to bimodal (${\rm V-I}$) color distributions seem to feature a fairly universal blue peak similar in color and dispersion to the one of the MW halo GC population (Larsen \etal 2001), while the second peak usually occurs at redder colors. GCs from the blue peak of the color distribution, as e.g. the old metal-poor halo GCs in the MW or M31, will be referred to as primary GCs in the following, while those from the red peak, as e.g. the more metal-rich disk or bulge GCs in the MW or M31, will be called secondary GCs. If the red-peak GCs in E/S0 galaxies are really younger and more metal-rich than the blue-peak GCs, however, still remains to be examined in detail. 

The apparent universality of the primary GC color distribution -- in terms of ${\rm \langle V-I \rangle ~ and ~ \sigma(V-I)}$ -- appearing in essentially all GC systems in very different types of galaxies (Larsen \etal 2001, Brodie \etal 2000) will be used as a basic assumption in our analysis. 

While there is agreement on the fact that any GC population is produced in some violent phase of SF, the precise circumstances are less clear. From observations of the ongoing or very recent formation of populous secondary GC systems in merging galaxies and merger remnants it seems clear that mergers of gas-rich galaxies can lead to the formation of new GC populations. Other scenarios, however, have as well been invoked, like e.g. {\sl in situ} two phase GC formation (Forbes \etal 1997) or hierarchical assembly (Cot\'e \etal 2000). 
To have color distributions in several bands, including UV and NIR, and successfully disentangle them into age and metallicity distributions may help to discriminate between the different suggestions.  

In a first step I use results from our new set of evolutionary synthesis models for Simple (= single burst) Stellar Populations ({\bf SSP}s) of various metallicities (Schulz \etal 2002) to explore the color evolution of star clusters of various metallicities in optical and NIR bands. I take the blue peak of the Milky Way  halo GC population with its mean metallicity ${\rm \langle [Fe/H] \rangle = -1.59}$ as a reference for the apparently universal primary population of GCs and analyse a large grid of arbitrary combinations of the two parameters age and metallicity for a theoretical secondary GC population to explore in which cases and at which ages this second population will be detectable as a second peak in the color distributions in the various bands and for which combinations of the parameters age and metallicity the two GC populations would appear hidden in one narrow or broadened color peak. 

Assuming a universal GC mass function (Ashman \etal 1995), evolutionary synthesis models also allow to study the luminosity functions ({\bf LF}s) of primary and secondary cluster populations in various bands in their time evolution. The LFs of -- putatively old --  GC systems are astrophysical standard candles with a very long range, out to more than Coma cluster distances ($\gta 100$ Mpc or ${\rm m-M \gta 35}$). Their turnover magnitudes seem to be fairly universal, ${\rm \langle M_{V_o} \rangle = -7.4 \pm 0.2}$, and are widely used for distance measurements out to 120 Mpc and determinations of the Hubble constant (see e.g. Kavelaars \etal 2000). GC LFs even look unimodal and inconspicuous in galaxies which clearly show bimodal GC color distributions (e.g. M87: Whitmore \etal 1995, NGC 1399: Geisler \& Forte 1990). 

In a second step I combine these results with the information we have about the redshift evolution of the average ISM abundances in spiral galaxies of various types. I will show that the broad age -- metallicity relation in the evolution of normal spiral galaxies acts to appreciably constrain the parameter space and only leaves color distributions and luminosity functions in agreement with current observations. 

The aim of the present study is to provide a multi-wavelength grid of color and luminosity distributions of any secondary GC population formed at some time in the past for comparison with observations.  

\section{Evolutionary Synthesis of GC Populations}
\subsection{SSPs of Various Metallicities}
Using Padova isochrones including the thermally pulsing AGB phase ({\bf TP-AGB}) and assuming a Salpeter IMF extending from 0.15 to 85 ${\rm M_{\odot}}$ we obtain the time evolution of spectra (90 \AA -- 160 ${\rm \mu m}$), colors, luminosities (U ... K), and stellar mass loss from ages $1 \cdot 10^8 - 15 \cdot 10^9$ yr for SSPs of 5 different metallicities [Fe/H] $=~-1.7,~-0.7,~-0.4,~0,~+0.4$ (Schulz \etal 2002). Inclusion of the TP-AGB phase in the stellar evolutionary tracks is very important for the (${\rm V-I}$) and (${\rm V-K}$) colors at ages $\gta 6 \cdot 10^7$ yr (see also Maraston \etal 2001), more so for higher metallicities than for lower ones. For ${\rm Z_{\odot}}$ it makes colors redder by $> 0.3$ mag in (${\rm V-I}$) and by $> 1$ mag in (${\rm V-K}$) at ages $1 \cdot 10^8 - 1 \cdot 10^9$ yr. This means that ages determined for young star clusters in several interacting galaxies or merger remnants from HST (${\rm V-I}$) colors and models without TP-AGB are significantly overestimated (cf. Schulz \etal 2002). 

Without going into details, we only recall here that -- compared to solar metallicity -- SSPs at lower metallicities are brighter in UBVRI and fainter in K, are bluer and have lower mass loss, lower M/L$_{\rm B,V}$ and higher M/L$_{\rm K}$ values throughout their evolution (Fritze -- v. A. 2000). Models are shown to not only correctly reproduce the observed GC colors after $12~-~13$ Gyr of evolution but also the empirical calibrations of colors vs metallicity, like (${\rm B-V}$) or (${\rm V-I}$) vs [Fe/H], for MW and M31 GCs. We showed that while for old clusters the color -- metallicity relations are fairly linear over the metallicity range of MW halo GCs ${\rm -2.3 \leq [Fe/H] \leq -0.5}$, they become significantly non-linear, however, towards metallicities ${\rm [Fe/H] > -0.5}$ and this holds true for Johnson/HST broad band colors as well as for Stroemgren ${\rm m_1}$, but less so for Washington ${\rm (C-T1)}$ (cf. Schulz \etal 2002, Fig. 9, 10). Clusters with ${\rm [Fe/H] > -0.5}$ will show colors significantly redder than those obtained from extrapolation of the empirical calibration. This implies that metallicities ${\rm [Fe/H] > -0.5}$ derived from colors using extrapolations of the empirical calibrations will be significantly overestimated. E.g. for an old GC with ${\rm (V-I_C)=1.2}$ the metallicity extrapolated from the observed calibration would be ${\rm \sim 3 \times Z_{\odot}}$, while in fact it only has solar metallicity, i.e. three times lower. Observationally, this theoretical result is supported by spectroscopy of individual GCs in NGC 1399 (Kissler -- Patig et al. 1999). 

Moreover, theoretical color--metallicity relations are shown to be significantly different for ages younger than 10 Gyr. Both effects, the age dependence and the non-linearity for [Fe/H] $>~ -0.5$, are important for the interpretation of the higher metallicity young and intermediate age star cluster populations observed in interacting galaxies, merger remnants, and dynamically young elliptical and S0 galaxies. 

\subsection{Modelling GC Populations}
\subsubsection{Color Distributions}
SSP models describe the time evolution of colors of individual star clusters of given metallicity. To simulate a cluster population, I assume that this cluster population -- similar to the MW halo GC system -- features a Gaussian distribution around the mean color appropriate for its metallicity and age. The width of the observed color distribution of MW and M31 halo GCs is determined by observational uncertainties and age differences to a small degree, and, largely, by the spread of individual GC metallicities around the mean values ${\rm \langle [Fe/H] \rangle^{MW~halo} = -1.59 ~and~ \langle [Fe/H] \rangle^{M31~halo} = -1.43 }$ (C\^ot\'e 1999, Barmby \etal 2000). This spread of metallicity is believed to reflect inhomogeneities in gas phase abundances at their birth. 

For simplicity and lack of better knowledge I describe the color distribution of a model cluster population of a given metallicity by a Gaussian around its mean color -- as given by SSP models for its metallicity and age -- with a dispersion identical to that observed for the MW and M31 GCs for that respective color (cf. Barmby \etal 2000, table 8). Thus, I implicitly assume that my model cluster populations have a scatter in metallicity comparable to that of the MW halo GCs and that the scatter induced on the colors has been constant during their evolution. Towards very young ages of order 1 to few Gyr, the color evolution of SSPs generally is less metallicity dependent. This means that the widths I assume for the color peaks of very young cluster populations are slightly overestimated.

The question, for which separation in color (or magnitude, see Sect. 2.2.2) two Gaussian-type GC populations may be resolved observationally depends on the number of clusters observed and on the observational accuracy. Ashman, Bird \& Zepf (1994) show that with sample sizes of 100 -- 200, typical of currently available data, a photometric precision of a factor 2.5 -- 3.0 less than the expected separation is required for a reliable detection or rejection of bimodality using the best available algorithms (KMM statistics and dip probability). For sample sizes as large as 500 -- 1000 objects, two peaks as close together as $2 \times$ the observational uncertainty may safely be separated. Larsen et al.'s (2001) data, e.g., have ${\rm \Delta(V-I) \lta 0.1~at~V_{lim} \sim 25~mag}$ for systems where bimodality is clearly detected. Assuming typical accuracies ${\rm \Delta(V-I) \lta 0.1}$ mag, the two peaks have to be separated by 0.25 -- 0.30 mag to be unambiguously resolved.

\subsubsection{Luminosity Functions}
For a given mass, SSP models also describe the luminosity evolution in terms of absolute luminosities or magnitudes of a star cluster of given metallicity. In addition to the small observational uncertainties and age and metallicity differences among halo GCs, the width of their LF is largely dominated by the cluster mass function. 
The mass function of MW GCs is log-normal with ${\rm \langle log (M_{GC}/M_{\odot}) \rangle = 5.47~ and ~ \sigma (log M_{GC}) = 0.50}$. For our purpose their LF is satisfactorily described by a Gaussian with ${\rm \langle M_V \rangle \sim -7.3~mag,~\sigma(M_V) \sim 1.3~mag}$ (Ashman \etal 1995). 
It is clear that secular destruction effects heavily act upon a GC population (e.g. Harris 1991). All clusters experience stellar evolutionary mass loss reducing their initial mass by 10 -- 20 \% over a Hubble time, depending on their IMF (Fritze -- v. A. 1998). Low mass clusters are prone to destruction by evaporation and tidal shocking. High mass clusters are preferentially destroyed by dynamical friction (e.g. Vesperini 1997, 1998, 2000, 2001). The mass function of a GC system at its birth unfortunately is still unknown. Luminosity (and hence M/L --) evolution is very strong in early stages (e.g. ${\rm \Delta M_V=2.5}$ to 2 mag between 4 Myr and 40 Myr for [Fe/H]$=-1.7$ and 0, respectively, and even more if gaseous emission contributions during the the earliest stages is included (cf. Anders \& Fritze - v. Alvensleben 2003). This strong early M/L-evolution acts to distort the shape of the LF with respect to that of the mass function through age spread effects in young GC systems observed in the process of formation, as e.g. in NGC 4038/39 (Meurer 1995, Meurer \etal 1995, Fritze -- v. A. 1999a). The ``true'' shape of its LF, and, hence, its mass spectrum are still a matter of debate (cf. e.g. Zhang \& Fall 1999, Fritze -- v. A. 1999a). General agreement, however, is reached on the fact that the strongest evolution of the mass function and intrinsic LF ($=$ referred to a uniform age) due to dynamical effects occurs within the first 1 -- 3 Gyr. 

We chose to assume a mass function identical in shape to the one observed for MW GCs for our model cluster systems, invariant in time for simplicity and lack of better knowledge. Hence, LFs of model GC populations are described by Gaussians with turn-over magnitudes ${\rm \langle M_{\lambda} \rangle,~\lambda = U~.~.~.~K}$, as given by our SSP models for a cluster of given age, metallicity, and a mass corresponding to ${\rm \langle log M \rangle _{GC}^{MW}}$. Widths ${\rm \sigma (M_{\lambda})}$ are assumed to be identical to those observed for M31 GCs in the respective wavelength bands $\lambda$ by Barmby \etal (2001). 

To visualise the color distributions and LFs of two theoretical GC populations, we further assume that both populations have comparable numbers of clusters. This, of course, need not be the case in any of the different GC formation scenarios mentioned above. Color distributions observed for a large number of galaxies (e.g. Gebhardt \& Kissler -- Patig 1999, Larsen \etal 2001, Kundu \& Whitmore 2001a, b) indeed contain examples of 2 GC color populations of different richness, although the numbers do not differ by large factors. Systems with two GC populations of vastly different richness -- if they exist -- will in any case be hard to detect observationally.

\subsection{Color Evolution of GC Populations}

\begin{figure*}[!htb]
\vspace{1.cm}
\epsfig{file=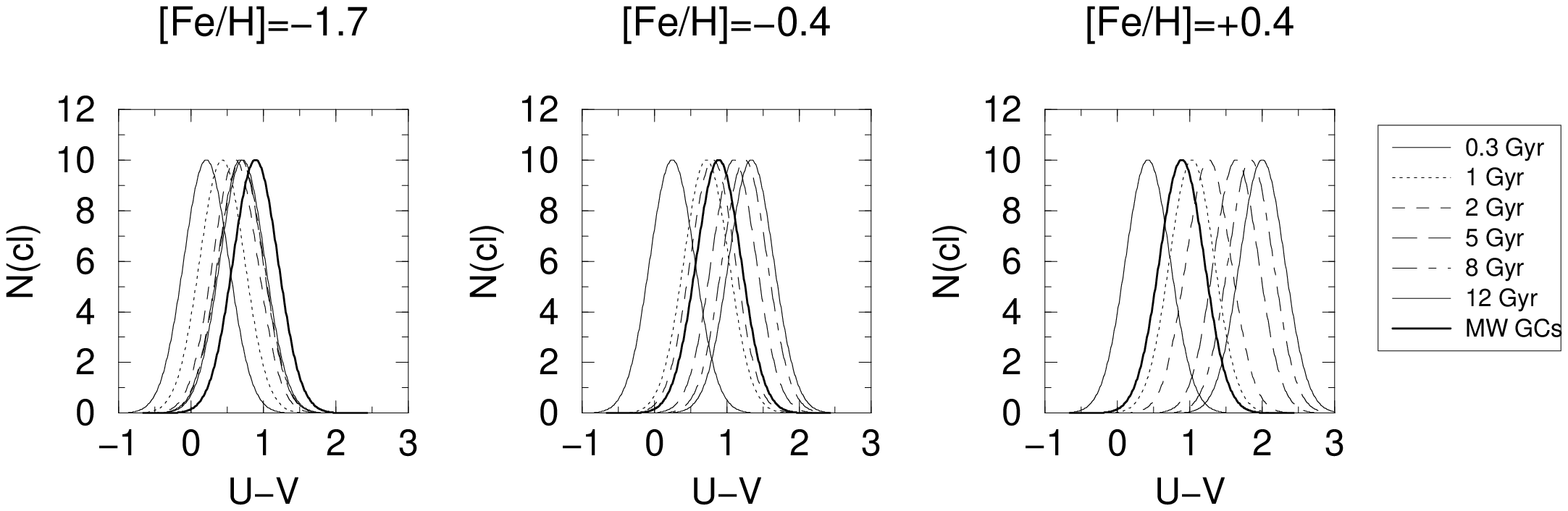,width=5.5cm,angle=270}
\hspace{-2.cm} \epsfig{file=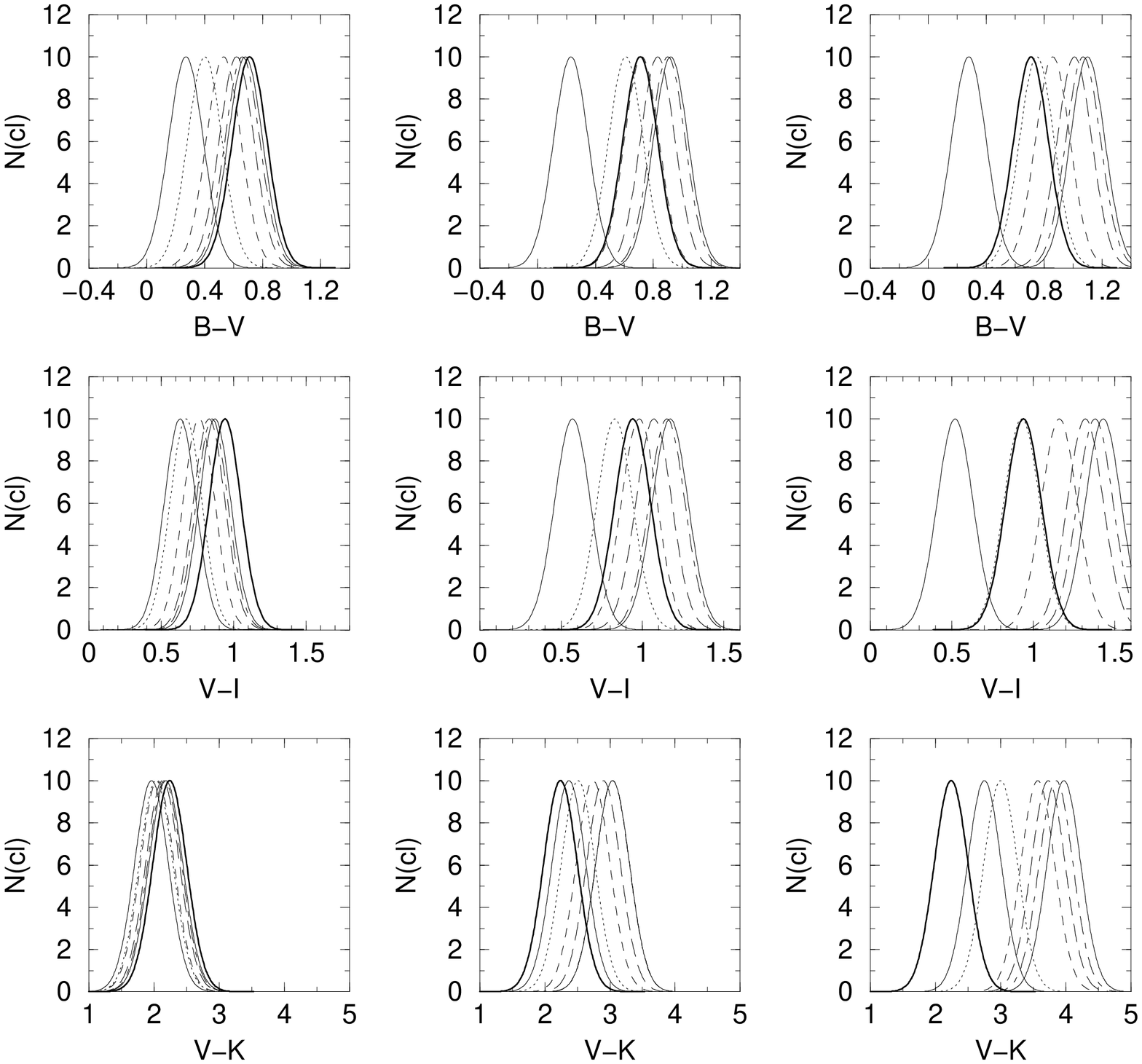,width=13.cm,angle=270}
  \caption{Color Evolution of a secondary GC population of various metallicities 
[Fe/H] $=-1.7$ (${\rm 1^{st}}$ column), [Fe/H] $=-0.4$ (${\rm 2^{nd}}$ column, 
[Fe/H] $=+0.4$ (${\rm 3^{rd}}$ column) in ${\rm (U-V),~(B-V),~(V-I)~and~(V-K)}$. 
The thick solid line shows the corresponding MW halo GC color distributions for 
reference.}
\end{figure*}

In our first step, we compare the time evolution of GC ($=$ SSP) model colors of various metallicities with the observed mean colors of MW halo GCs from Barmby et al.'s (2000) compilation. 

Fig. 1 presents the time evolution of color distributions ${\rm U-V}$, ${\rm B-V}$, ${\rm V-I}$, and ${\rm V-K}$ for SSPs of various metallicities ${\rm [Fe/H] = -1.7}$ (${\rm 1^{st}}$ column), ${\rm [Fe/H] = -0.4}$ (${\rm 2^{nd}}$ column), and ${\rm [Fe/H] = +0.4}$ (${\rm 3^{rd}}$ column) compared to the corresponding observed MW halo GC color distributions. 

We use the mean colors and color dispersions for MW GCs as given by Barmby \etal (2000) ${\rm \langle U-V \rangle = 0.89}$, ${\rm \sigma (U-V)=0.31}$, ${\rm \langle B-V \rangle = 0.71}$, ${\rm \sigma (B-V)=0.12}$, ${\rm \langle V-I \rangle = 0.94}$, ${\rm \sigma (V-I) = 0.11}$, ${\rm \langle V-K \rangle = 2.24}$, ${\rm \sigma (V-K) = 0.26}$. The normalisation of the maximum number of clusters to ${\rm N_{cl}=10 ~ at ~ \langle U-V \rangle, ~\langle B-V \rangle}$, ..., respectively, is arbitrary but the same for all cases. We construct cluster population models assuming that they have the same color dispersions as the observed ones throughout their evolution. 

Fig. 1 shows how model cluster populations move through the different color diagrams as they age from 0.3 to 12 Gyr. 
In the optical colors ${\rm U-V}$, ${\rm B-V}$, and ${\rm V-I}$, very low metallicity clusters (${\rm [Fe/H] = -1.7}$) do not quite reach those of MW halo GCs ($\rm \langle [Fe/H] \rangle = -1.59$ cf. C\^ot\'e 1999) at 12 Gyr, although they come very close in the later stages of their evolution. 
Intermediate metallicity clusters (${\rm [Fe/H] = -0.4}$) also start out very blue but evolve rapidly towards MW GC colors within only 1 -- 2 Gyr. They show redder colors than MW GCs at ages $\geq 5$ Gyr. 
Metal-rich model clusters (${\rm [Fe/H] = +0.4}$) already show MW GC colors at ages around 1 Gyr and become significantly redder later-on. 
In ${\rm V-K}$, only the ${\rm [Fe/H] = -1.7}$ clusters start out slightly bluer than today's MW halo GCs, almost reaching their colors in late evolutionary phases. At ${\rm [Fe/H] = -0.4}$, they are already redder than MW GCs at ages $\sim 3$ Gyr and go on reddening thereafter to ${\rm V-K = 3}$ at 12 Gyr. 
At high metallicity ${\rm [Fe/H] = +0.4}$ model clusters are already redder than old MW halo GCs by $\sim 0.5$ mag at ages $\sim 0.3$ Gyr. Around 12 Gyr, they will reach ${\rm V-K \sim 4}$. 

Note that certain combinations in age and metallicity, that are indistinguishable in optical colors due to the well-known age -- metallicity degeneracy, do split up in (${\rm V-K}$). This is seen e.g. for clusters with [Fe/H]$= -0.4$ and ages around 5 Gyr or clusters with [Fe/H]$=+0.4$ and ages around 1 Gyr. For very low metallicity clusters [Fe/H]$=-1.7$ the opposite is true. They show very uniform ${\rm (V-K)}$ values at all ages while splitting up in optical colors.   

We draw the following {\bf conclusions} from this part of our analysis: {\bf 1.)} An observed unimodal GC color distribution need not necessarily imply that all GCs have the same age and metallicity. Two (or more) GC populations may hide within one peak. {\bf 2.)} In case of an observed bimodal GC color distribution it is not {\sl a priori} clear if a red or blue peak is due to an older/younger or metal-rich/metal-poor cluster population. Observations in the NIR help to decide in some cases, individual cluster spectroscopy may be required for particularly tricky ones.

\subsection{Luminosity Evolution of GC Populations}

\begin{figure*}[!htb]
\vspace{0.5cm}
\epsfig{file=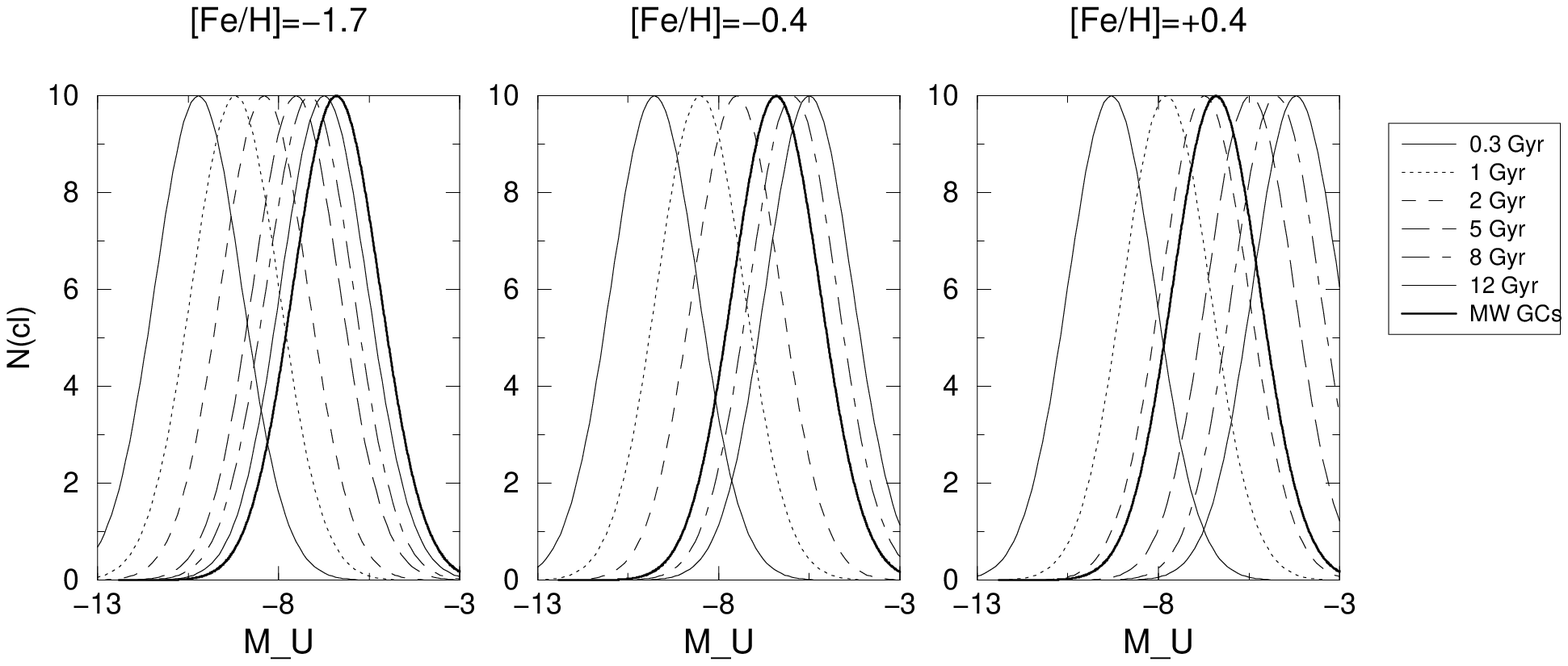,width=6.2cm,angle=270}
\hspace{1.truecm}\epsfig{file=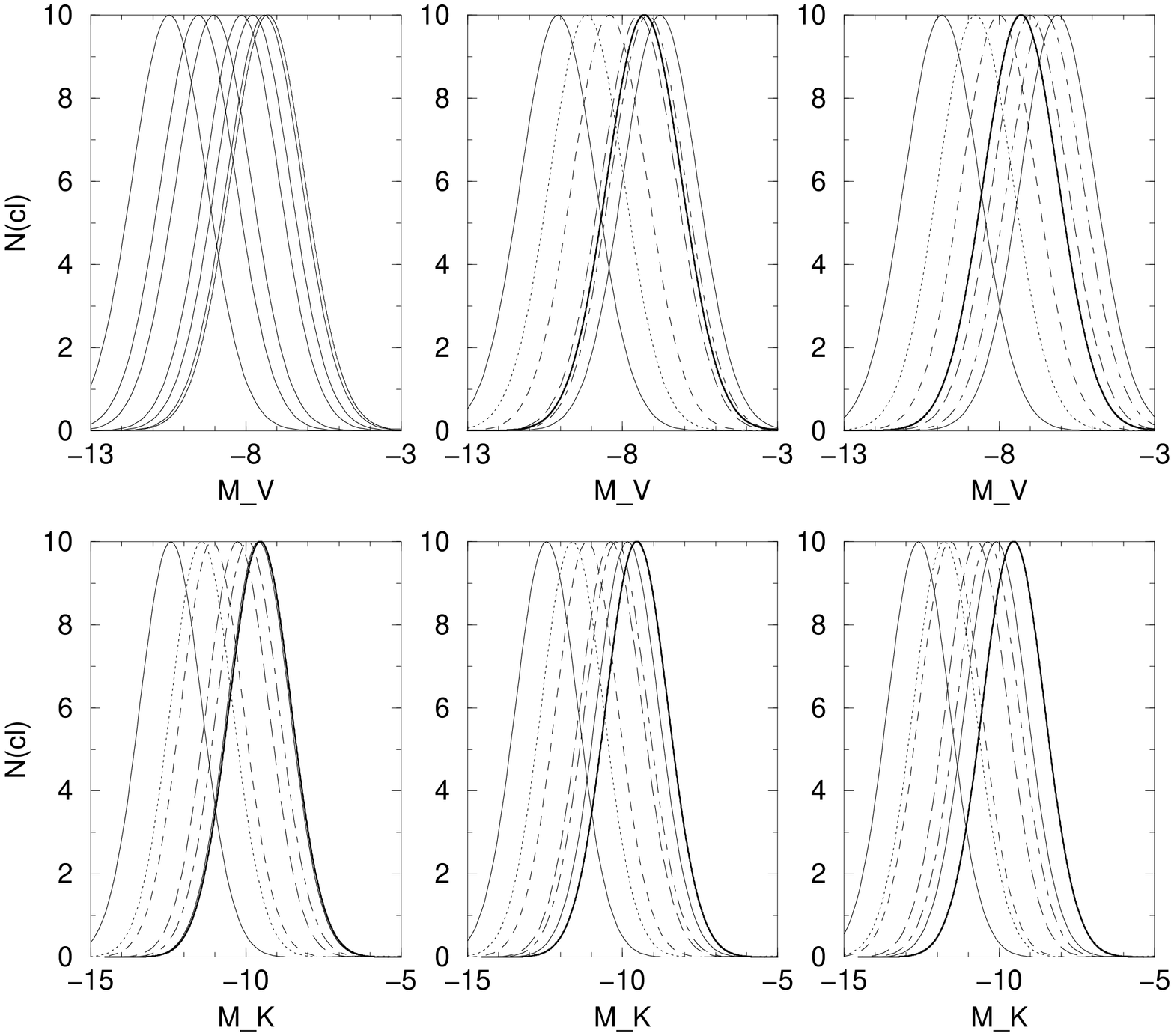,width=11.cm,angle=270}
  \caption{Luminosity Evolution of a secondary GC population of various metallicities 
[Fe/H] $=-1.7$ (${\rm 1^{st}}$ column), [Fe/H] $=-0.4$ (${\rm 2^{nd}}$ column, 
[Fe/H] $=+0.4$ (${\rm 3^{rd}}$ column) in ${\rm M_U}$ (top panels), ${\rm M_V}$ (middle panels) and ${\rm M_K}$ (bottom panels). 
The thick solid line shows the corresponding MW GC LFs for 
reference.}
\end{figure*}

The dispersion of the observed MW GC color distribution is a result of observational uncertainties and of the scatter in metallicity around the mean ${\rm \langle [Fe/H]\rangle}$ of the MW clusters. The width of the Gaussian well approximating the MW GC LF is also affected by these two effects but certainly dominated by the mass function of MW GCs. 

SSP models give the evolution of absolute luminosities ${\rm M_{UBVRIK}}$ as a function of the total mass initially turned into stars. 
To study the time evolution of cluster LFs at various metallicities, we assume a mass function similar to that of the MW GCs (Ashman \etal 1995) and gauge the luminosity ${\rm M_V}$ of an SSP with [Fe/H]$=-1.34$, the mean metallicity of all MW GCs, at an age of 12 Gyr to the observed MW GC turn-over ${\rm \langle M_{V_o}\rangle = -7.3}$ mag that corresponds to a mean GC mass of ${\rm \sim 3 \cdot 10^5~M_{\odot}}$ (Ashman \& Zepf 1998). 

Fig. 2 shows the observed MW GC LFs in three different wavelength bands U, V, and K together with the time evolution of our model cluster LFs for three different metallicities. The U- and K-band LFs for MW GCs are obtained from that in the V-band by a combination of ${\rm \langle M_V \rangle = -7.3}$ with the observed ${\rm \langle U-V \rangle = 0.89}$ and ${\rm \langle V-K \rangle = 2.24}$ which give ${\rm \langle M_U \rangle = -6.4}$ and ${\rm \langle M_K \rangle = -9.5}$ mag. Barmby \etal (2001) analyse M31 halo GC LFs in the wavelength bands U, B, V, R, J, and K and show that the dispersions ${\rm \sigma (M_{U_o})}$, ${\rm \sigma (M_{V_o})}$, and  ${\rm \sigma (M_{K_o})}$ are identical within the uncertainties and that, indeed, ${\rm \langle U-V \rangle -  (\langle M_{U_o} \rangle - \langle M_{V_o} \rangle) = 0.02}$ and ${\rm \langle V-K \rangle -  (\langle M_{V_o} \rangle - \langle M_{K_o} \rangle) = 0.04}$.  

Low metallicity model cluster populations with ${\rm [Fe/H] = -1.7}$ start out very bright early in their evolution with ${\rm \langle M_U \rangle \sim -10.2,~ \langle M_V \rangle \sim -10.5,~and~ \langle M_K \rangle \sim -12.4}$ at an age of 0.3 Gyr, fade at a rate decreasing in time and reach the MW GCs' ${\rm \langle M_{U_o} \rangle}$, ${\rm \langle M_{V_o} \rangle}$, and ${\rm \langle M_{K_o} \rangle}$ around 12 Gyr. Intermediate metallicity cluster populations with ${\rm [Fe/H] = -0.4}$ show a very similar evolution in early stages $\leq 2$ Gyr in U, V, and K to the low metallicity ones. At an age of $\sim 5$ Gyr they have reached the MW GCs' ${\rm \langle M_{U_o} \rangle}$ and ${\rm \langle M_{V_o} \rangle}$ and fade slightly more later on. At an age of 12 Gyr, intermediate metallicity clusters will be fainter by $\sim 1$ and 0.5 mag in ${\rm \langle M_{U} \rangle}$ and ${\rm \langle M_{V} \rangle}$, respectively, than MW GCs while in ${\rm \langle M_K \rangle }$, they remain slightly brighter at 12 Gyr than MW GCs. 

At high metallicity ${\rm [Fe/H] = +0.4}$, model cluster populations become less luminous in ${\rm \langle M_U \rangle }$ and ${\rm \langle M_V \rangle }$ than MW GCs already at ages of 8 and 5 Gyr, respectively. At 12 Gyr they are fainter than MW GCs by more than 2 and 1 mag, respectively,  in U and V. Again, in ${\rm \langle M_K \rangle}$ they fade less than the lower metallicity clusters and end up $\sim 0.7$ mag brighter in ${\rm \langle M_K \rangle}$ than MW GCs.

\subsection{Comparing Color Distributions and LFs}
Comparing color distributions and LFs of two theoretical cluster populations, all possibilities are seen to appear depending on the specific combination of ages and metallicities. 
We choose to discuss this for cluster populations of various metallicities and ages in comparison to the observed color distributions and LFs of MW GCs, our reference system. 

Very low metallicity clusters at young ages show both color distributions and LFs distinctly different from those of MW GCs. 
Intermediate metallicity clusters at young ages, around 1 Gyr, show color distributions in ${\rm U-V}$, ${\rm B-V}$, ${\rm V-I}$, and ${\rm V-K}$ that do not differ significantly from those of MW GCs while their LFs, in U, V and K, are all clearly different from their MW analogues. They peak at brighter ${\rm \langle M_U \rangle,~ \langle M_V \rangle ~ and ~ \langle M_K \rangle}$. 
At older ages of 8 -- 12 Gyr, however, such an intermediate metallicity cluster population shows distributions in all colors significantly different from those of the MW GCs while their LFs now are very similar to those of the MW GCs in V and K, but slightly fainter in U. 
High metallicity clusters at a young age of $\sim 1$ Gyr show ${\rm U-V}$, ${\rm B-V}$ and ${\rm V-I}$ distributions fairly similar to those of old MW GCs, while their ${\rm V-K}$ distribution is already significantly redder. Their LFs in U, V, and even more so in K, still peak at much brighter magnitudes. Their ${\rm U-V}$, ${\rm B-V}$, and ${\rm V-I}$ colors start to become significantly redder than those of MW GCs at ${\rm t \gta 2}$ Gyr. Around 2 Gyr a clear difference to MW GCs is still seen in ${\rm \langle M_K \rangle}$, but no longer in ${\rm \langle M_V \rangle }$ or ${\rm \langle M_V \rangle }$. 

So for two theoretical GC populations, an old and moderately metal-poor one like those of the MW or M31 halos and a second one with an arbitrary combination of age and metallicity, all situations are possible: bimodal color distributions and bimodal LFs, single-peak color distributions and, at the same time, bimodal LFs, single-peak color distributions and single-peak LFs, and bimodal color distributions together with single-peak LFs. 

If this were true in reality, it would put into question the method of distance determination using GC LFs. It seems, however, to be at variance with the observational situation where even in systems with clearly bimodal GC color distributions like M87 or NGC 4472, the LFs look unimodal and fairly normal. We will come back to this issue and its implications for the range of allowed combinations of metallicities and ages in Sect. 5.3. 

\section{Spirals-Spiral Mergers at Various Redshifts and the Formation of GCs}

\subsection{Redshift Evolution of Undisturbed Galaxies}
Evolutionary synthesis models para\-me\-trise galaxies of various spectral types by their respective star formation ({\bf SF}) histories. Starting from an initial gas cloud and specifying a SF history and an IMF, our models simultaneously yield the spectrophotometric evolution of the stellar population and the chemical evolution in terms of ISM abundances, both as a function of time and, for any kind of cosmological model, also as a function of redshift. Following both aspects of galaxy evolution simultaneously allows us to describe the evolution in a chemically consistent way, i.e. to monitor for every individual star its initial metallicity as given by the gas phase abundance at birth, and to follow each star  on a stellar evolutionary track and with yields appropriate for its metallicity (Fritze -- v. A. 1999b). This produces results in good agreement with observed stellar metallicity distributions, template spectra from UV through optical, colors from U through K, and characteristic HII region abundances of local galaxy types as well as with the observed redshift evolution of luminosities and colors (M\"oller \etal 1999). Comparison with precise individual element abundances measured from high resolution spectra of Damped Ly$\alpha$ absorbers for a series of elements (Fe, Si, Zn, Cr, Ni, S, Al, Mn) shows good agreement with the redshift evolution of the absolute abundances in our spiral models over a redshift range from ${\rm z>4}$ to ${\rm z\sim 0.4}$. By ${\rm z=0}$, models directly match observed characteristic HII region abundances (cf. Lindner \etal 1999 for details). The current discussion if perhaps not all Damped Ly$\alpha$ absorbers are the progenitors of today's spiral galaxies does not affect our conclusions concerning ages and metallicities of secondary GCs as long as our continuous SF histories remain valid approximations to the average SF histories of (all the building blocks of) the respective spiral types. In any case, the requirement for models to simultaneously match both the average spectral and chemical properties of specific galaxy types (E, S0, Sa, . . ., Sd) allows for significantly tighter constraints on the characteristic SF histories than either aspect alone and leads us to believe that we reasonably understand the redshift evolution of the average ISM abundances in spiral galaxies over significant lookback times. 

Note, however, that our models at this stage still are simplified global 1-zone closed-box models without any spatial resolution nor dynamics and only consider one gas phase. While the finite and metallicity-dependent stellar lifetimes are fully taken into account, we assume instantaneous and perfect mixing of the recycled enriched gas into the reservoir.

\subsection{GC Formation in Mergers}
Numerous observational examples are known where in the powerful starbursts accompanying mergers of gas-rich spirals populous systems of young star clusters are born (NGC 7252, NGC 3921, NGC 4038/39, ...). Many, if not most, of these compact bright young star clusters probably are young GCs (cf. Schweizer 2002 for a recent review, Carlson \& Holtzman 2001). Based on the detailed analysis of the starburst and the properties of the progenitor galaxies in NGC 7252 we predicted abundances for the young clusters in this Sc -- Sc merger remnant of ${\rm \gta \frac{1}{2}~Z_{\odot}}$ (Fritze -- v. A. \& Gerhard 1994b). Spectroscopy of the brightest clusters, i.e. those formed towards the end of the extended starburst, yielded abundances ${\rm Z \sim Z_{\odot}}$ with some indication of $\alpha$-element enhancement from the burst itself (Schweizer \& Seitzer 1993, 1998, Fritze -- v. A. \& Burkert 1995, Maraston \etal 2001) in good agreement with model predictions. Note that the strong burst that largely consumed the available gas in the inner part of this post-burst galaxy  rapidly increased the ISM metallicity (cf. Fig. 12 in Fritze -- v. A. \& Gerhard 1994a). 

Together with the primary GC populations inherited from the merging galaxies, these secondary GC populations formed in strong starbursts should be detectable around merger remnants. If not hidden within the color distribution of the primary one by a conspiracy between its younger age and higher metallicity, such a secondary GC population may testify to a merger/starburst origin of its parent galaxy, even at times when most of the other tracers like tidal features, blue colors, Balmer absorption lines, fine structure, etc. will have disappeared.

In the following, I will use the redshift evolution of the ISM in our spiral galaxy models to investigate the metallicities of theoretical secondary populations of GCs forming from the ISM in spiral -- spiral mergers at various redshifts. Combining this with SSP models for the respective metallicities, the time evolution of colors and characteristic luminosities of these secondary GCs are obtained. For simplicity I will restrict myself to mergers among equal type galaxies, i.e., Sa -- Sa, . . ., Sd -- Sd.

\begin{figure}
  \centering \epsfig{file=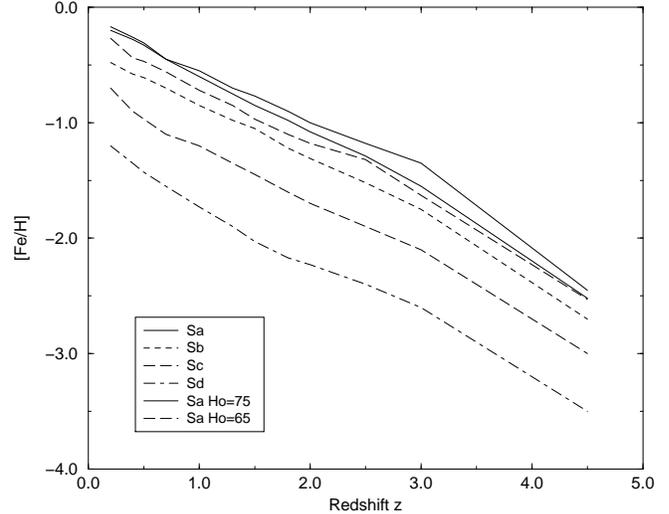,width=7.cm,angle=270}
  \caption{Redshift evolution of average ISM abundances for different spiral models Sa, Sb, Sc, Sd and ${\rm (H_o,~\Omega_o,~\Omega_{\Lambda})~=~(50,~1.0,~0)}$. Thin lines show the Sa model in two other cosmologies: ${\rm (H_o,~\Omega_o,~\Omega_{\Lambda})~=~(75,~0.1,~0)}$ and ${\rm (65,~0.3,~0.7)}$.}
\end{figure}

The redshift evolution of ISM abundances for various spiral types, as presented in Fig. 3 for the example of [Fe/H], shows that only a limited range of combinations of metallicity and age for secondary GCs is expected to result from mergers of normal spirals. Despite a significant amount of scatter at fixed redshift, introduced by the metallicity differences among different spiral types, a broad cosmic age -- metallicity relation is observed (cf. Fritze -- v. A. 2001). The ISM metallicity in spirals increases slowly with decreasing redshift. In Lindner \etal (1999) we show, that, indeed, our models bracket the range of Damped Ly$\alpha$ abundance obervations, not only for iron, but for all of the 8 elements (Fe, Zn, Si, Cr, Ni, Al, S, Mn) for which reasonable numbers of abundance determinations exist, over the redshift range ${\rm 0.4 \leq z \leq 4.4}$ and match the characteristic ISM abundances of local spirals at ${\rm z = 0}$. Also seen in Fig. 3 is the effect of a change in the cosmological parameters on the redshift evolution of the ISM abundances. The transformation from time to redshift, of course, depends on the cosmological parameters adopted, but among cosmologies currently discussed, like ${\rm (H_o=50,~\Omega_o=1,~\Omega_{\Lambda}=0),~(H_o=75,~\Omega_o=0.1,~\Omega_{\Lambda}=0)}$, and ${\rm (H_o=65,~\Omega_o=0.3,~\Omega_{\Lambda}=0.7)}$, the changes are small, much smaller in fact than the differences among neighboring spiral types.

Hence, secondary GCs formed out of this gas during merger-induced starbursts are predicted to be the more metal-rich the later the merger occurs, i.e. the younger the newly formed GCs are at the present time, and the earlier the spectral type of the merging galaxies.  
Note that relaxing the closed-box approximation and, e.g., admitting for infall of primordial gas at a rate either constant or continuously decreasing with decreasing redshift does not drastically affect our results. Once SFRs are readjusted to again give agreement with present-day galaxy properties, the broad cosmic age -- metallicity relation is essentially recovered. 

In our closed-box models, the gas content is meant to describe the average total gas content in the corresponding galaxy type, i.e. to include the HI observed beyond the optical radius of spiral disks. Nevertheless and in agreement with observations it decreases to low values as early-type galaxies reach ISM abundances ${\rm \gta (\frac{1}{2} - 1) \cdot Z_{\odot}}$. If these early-type galaxies merged not too long ago, no strong bursts and not much secondary cluster formation are expected due to the shortage of gas supply for SF. Galaxies of a given type may feature a wide range of gas-to-stellar mass ratios and our models reproduce typical values. The amount of gas beyond (and sometimes far beyond) the optical radius, in particular, is not very well known yet for representative samples of galaxies. How much of this gas may move inward, how much of it may be involved in the starburst, and what is the ratio of field star to cluster star formation may depend on details of the merger. We therefore refrain from using the model gas content as a boundary condition for the number of secondary GC that can be formed in mergers. The question, how many of those secondary GCs can survive for how long in the environment of a merger is another open issue.

SSP models have shown that metallicity effects on the evolution of a GC depend both on wavelength/color and on age. Hence, it is the interplay of age and metallicity that determines the color and luminosity evolution of a secondary GC population in a way that is specific for each wavelength region.

\section{Colors and Luminosities of Secondary GCs Formed in Spiral-Spiral Mergers}
In this Section, we present the color and luminosity evolution of secondary GC populations  forming in spiral -- spiral mergers at various redshifts. We combine the redshift evolution of spiral ISM abundances with SSP models for the color and luminosity evolution of star clusters of various metallicities. We have to know the ages that these clusters will have obtained by ${\rm z = 0}$ to predict their present-day colors and luminosities for comparison with observations. The relation between the time from cluster formation at some redshift z until today depends on the choice of the cosmological parameters ${\rm H_o,~\Omega_o,~and~\Lambda_o}$ and, at high redshift to some degree on the assumed redshift of galaxy formation ${\rm z_f}$, as well.

\begin{table}[htbp]
\begin{center}
\caption{Present age of secondary star clusters forming in mergers at different redshift for 3 different cosmologies ${\rm (H_o,~\Omega_o,~\Omega_{\Lambda})~=~(50,~1.0,~0),~(75,~0.1,~0),~(65,~0.3,~0.7)}$, respectively, and a redshift of galaxy formation ${\rm z_f \geq 5}$.}
\begin{tabular}{|c|c|c|c|}
\hline
 & & & \\ 
 Redshift & GC Age [Gyr] & GC Age [Gyr] & GC Age [Gyr] \\
 & & & \\
 z & ${\rm (50,~1.0,~0)}$ & ${\rm (75,~0.1,~0)}$ & ${\rm (65,~0.3,~0.7)}$ \\
 & & & \\
\hline
 & & & \\
 4.0 & 11.9 & 10.0 & 13.2 \\
 3.0 & 11.4 &  9.4 & 12.6 \\
 2.0 & 10.5 &  8.4 & 11.4 \\
 1.5 &  9.7 &  7.6 & 10.3 \\
 1.0 &  8.4 &  6.4 &  8.6 \\
 0.5 &  5.9 &  4.3 &  5.6 \\
 0.2 &  3.1 &  2.2 &  2.7 \\
 & & & \\
\hline

\end{tabular}
\end{center}
\end{table}

Table 1 shows the dependence of the relation between the redshift of formation of secondary GCs and their present-day age on the choice of the cosmological parameters within their currently discussed ranges. Among models with ${\rm (H_o=50,\:\Omega_o=1,\:\Lambda_o=0)}$, ${\rm (H_o=75,\:\Omega_o=0.1,\:\Lambda_o=0)}$, and ${\rm (H_o=65,\:\Omega_o=0.3,\:\Lambda_o=0.7)}$ age differences at fixed redshift rarely reach 25 \% and are much smaller at most redshifts. Hence, our results are not strongly dependent on the specific cosmological model as long as it allows for an age of the old GC population of about 12 Gyr. 

\subsection{Color Evolution}

In Fig. 4 we present the colors in ${\rm (B-V)}$, ${\rm (V-I)}$, and ${\rm (V-K)}$ that an average secondary GC formed in spiral-spiral mergers at various times/redshifts will have attained by today. As we have seen in Sect. 2.3. that ${\rm (U-V)}$ color distributions do not give much more information beyond that contained in ${\rm (B-V)}$ we do not discuss ${\rm (U-V)}$ here any further.

We also plot in Fig. 4 the mean colors together with their 1 $\sigma$ ranges of the Milky Way halo GC system which we quoted as a reference for the universal primary GC population to explore in which cases secondary GC populations formed in a variety of galaxy mergers may be discriminated from the primary one pre-existing in the merging spirals and assumed to be similar to the blue GC populations in the MW or M31. We recall that the abundances of our merging spiral models are average abundances of the respective galaxy types as measured at 1 ${\rm R_{eff}}$. In case of particularly bright spirals, cluster formation constrained to very central regions, or self-enrichment during an extended starburst, GC abundances may be somewhat higher and colors hence somewhat redder, in case of lower than average luminosity spirals or if formed out of material from beyond ${\rm \sim 1~ R_{eff}}$ secondary GCs may have slightly lower metallicity and hence bluer colors than our simplified models predict. 

GCs formed in an Sa -- Sa merger at ${\rm z \sim 0.2}$, i.e. only $\sim 3$ Gyr ago, are predicted by our models to have a metallicity ${\rm [Fe/H] \sim -0.2}$ and, hence, to have ${\rm (B-V) \sim 0.86}$. Until ${\rm z \sim 1}$ the higher ages that clusters then will have achieved by today are balanced by their lower metallicities to keep ${\rm (B-V) \sim constant}$. For clusters formed at ${\rm z > 1}$ the metallicity decrease wins against the age increase in terms of the ${\rm (B-V)}$ evolution and clusters get bluer. GCs forming in mergers of the significantly less metal-enriched Sd -- Sd mergers will have much bluer colors, around ${\rm (B-V) \sim 0.65}$, if they form later than ${\rm z \sim 1.3}$ and even bluer if they form earlier than that. 

It is seen that for most types of galaxy mergers at ${\rm z \lta 1.5}$ the predicted ${\rm (B-V)}$ color falls within the color range of the primary population. Only for mergers (of preferentially later spiral types) at redshifts ${\rm z \gta 2.5}$ secondary GCs are expected to be significantly bluer in ${\rm (B-V)}$ than MW halo GCs. The low gas content of early-type spirals at ${\rm z < 1}$ will probably preclude the formation of important populations of GCs significantly redder today in ${\rm (B-V)}$ than MW GCs.

We conclude that {\bf in most cases it will be impossible to identify a secondary GC population formed in spiral -- spiral mergers from ${\rm B-V}$ color distributions of GCs associated with their merger remnants.}

\begin{figure}[!htb]
\centerline{\vbox{
\psfig{figure=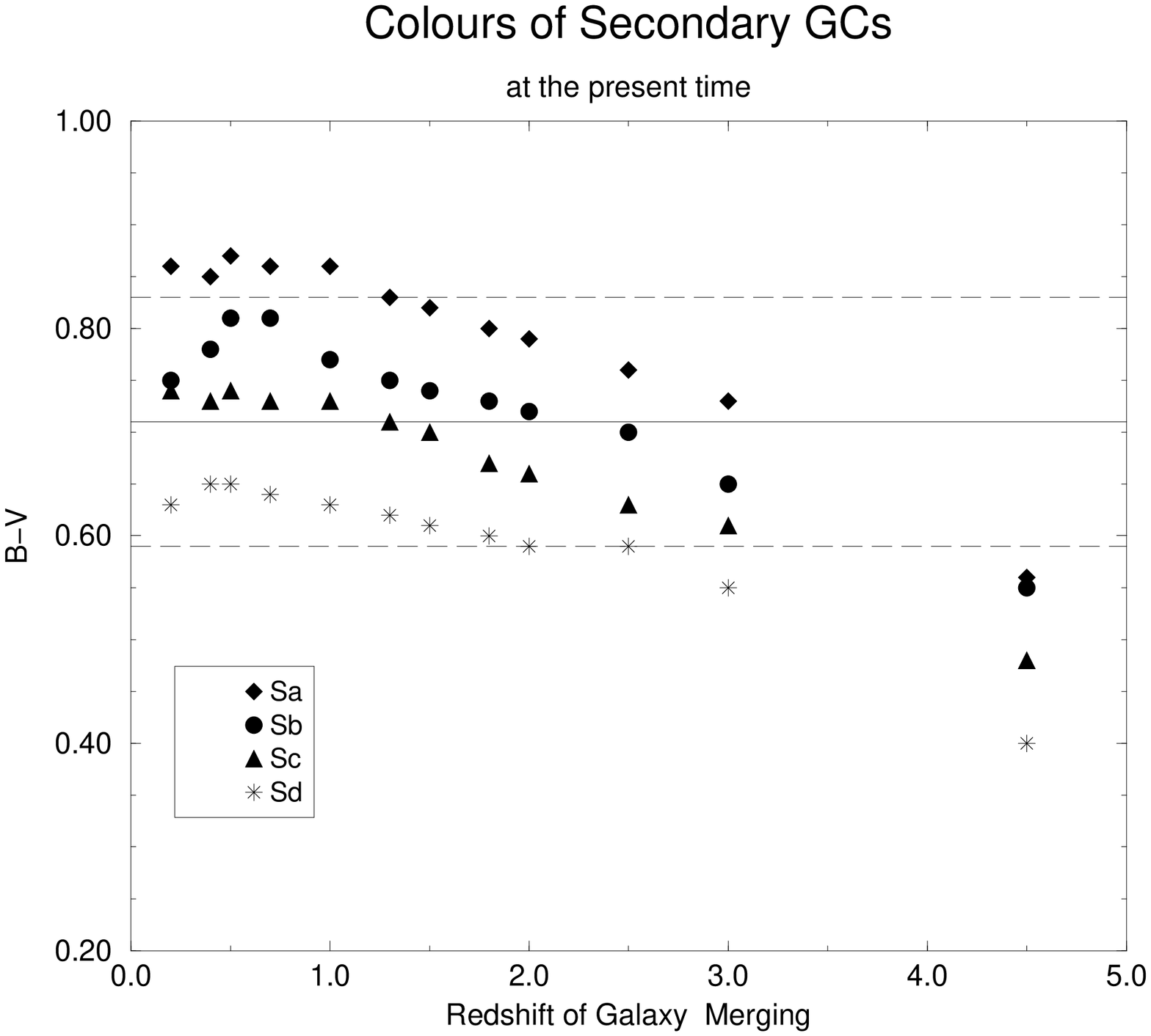,width=6.7cm,angle=270}
\psfig{figure=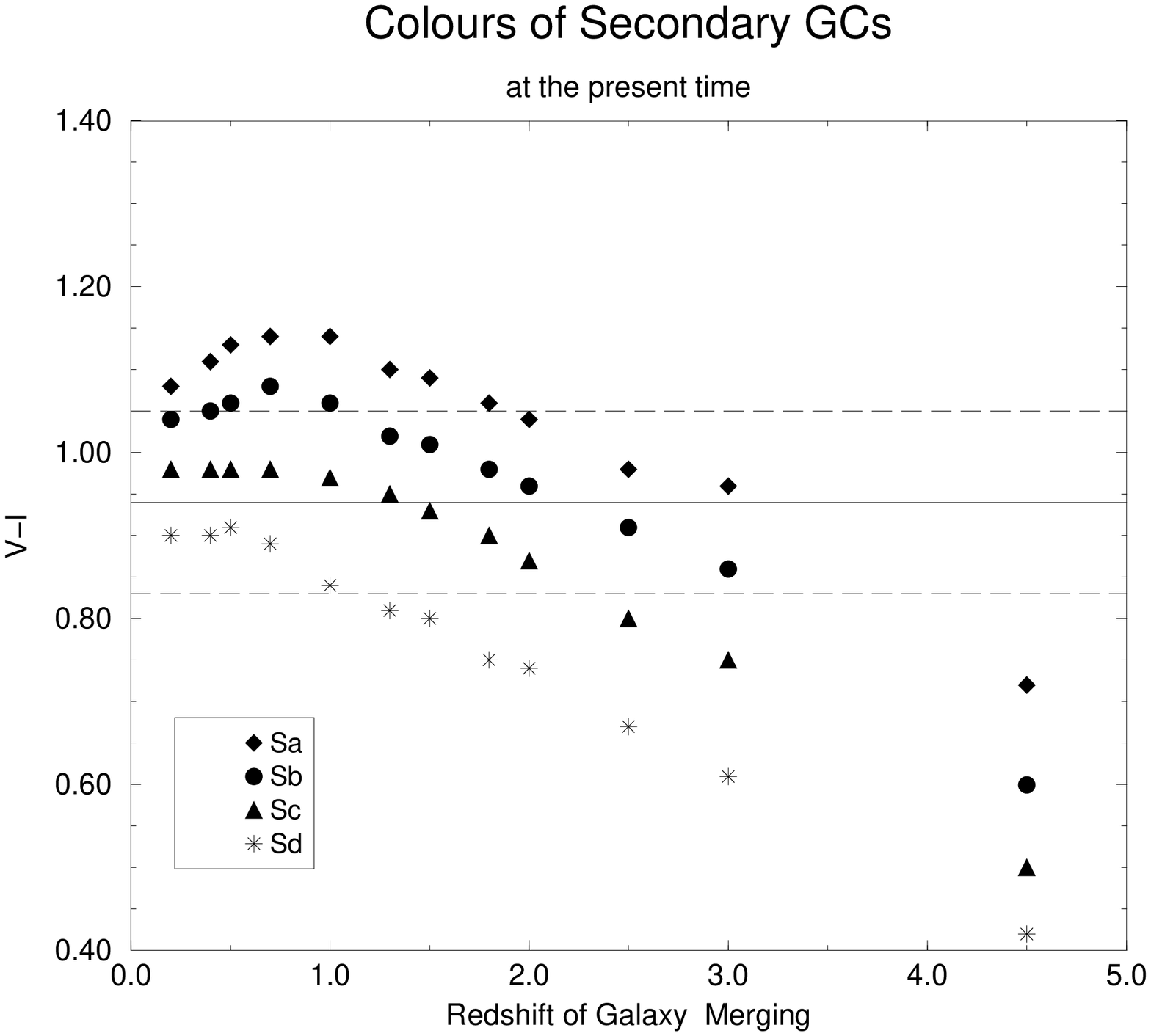,width=6.7cm,angle=270}
\psfig{figure=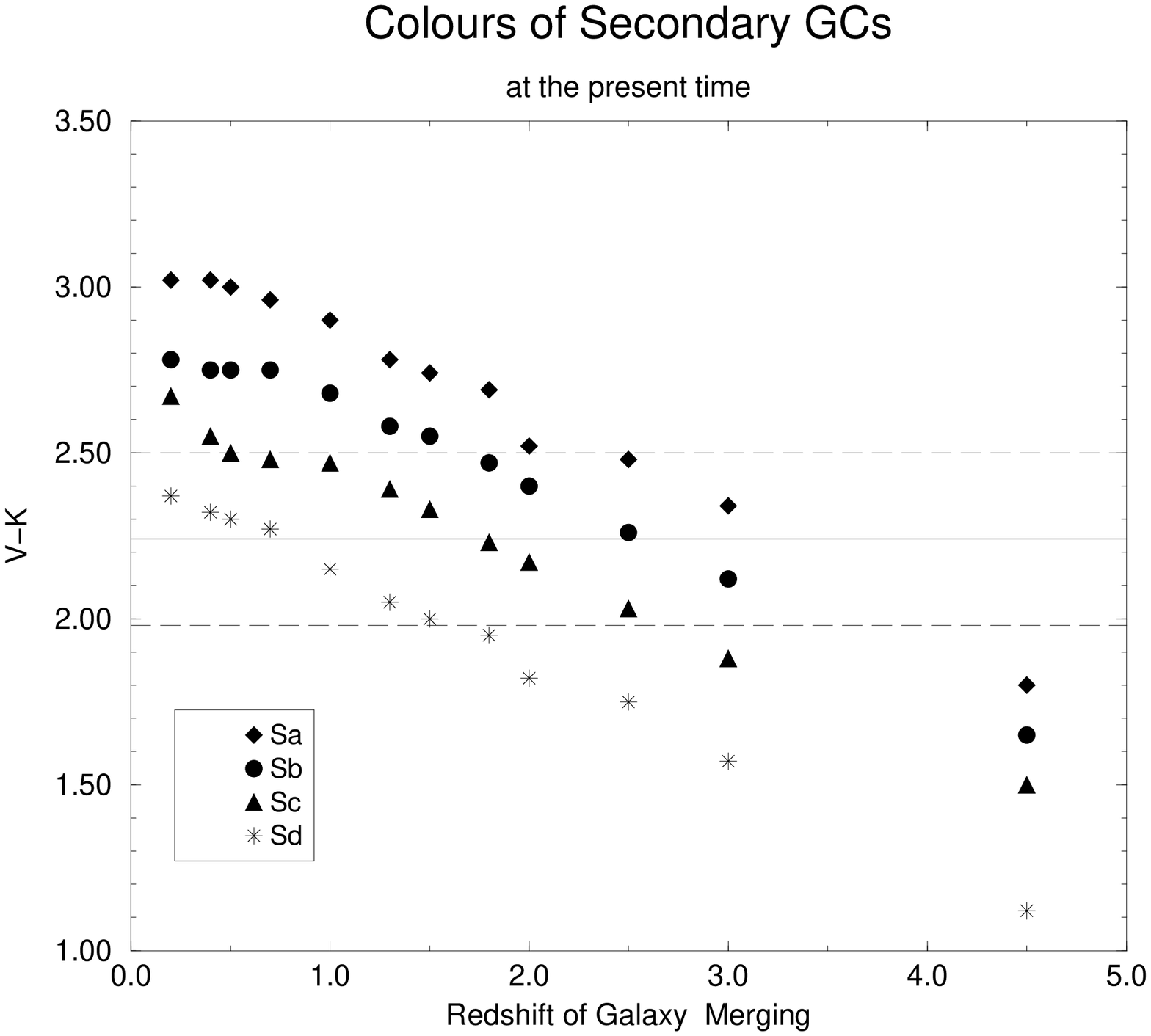,width=6.7cm,angle=270}
}}
\caption{Present day colors ${\rm (B-V,~(V-I),~and~(V-K)}$ of secondary GCs formed in mergers of various galaxy types occuring at different redshifts. Horizontal lines mark mean colors of MW GCs (solid) and their 1 $\sigma$ ranges (dashed lines). For individual GCs in M31, e.g., observational color uncertainties as given by Barmby \etal (2000) are $\pm 0.01$ in ${\rm (B-V)}$ and ${\rm (V-I)}$ and $\pm 0.02$ in ${\rm (V-K)}$, comparable to the size of our symbols.}
\end{figure} 

Comparison with the results for ${\rm (V-I)}$ and ${\rm (V-K)}$ colors reveals considerable differences. In those colors, there are many more cases in which the color distributions of GCs associated with a merger remnant are expected to reveal the secondary GC population. 

In ${\rm (V-I)}$, relatively late mergers of early-type spirals at ${\rm z \gta 0.5 - 1}$, i.e. before their gas content gets too low, may well produce GC populations with present-day colors {\bf redder} than those of the primary population. GCs forming in late-type mergers at ${\rm z \gta 1 - 2}$ are predicted to be {\bf bluer} in ${\rm (V-I)}$ than the primary ones. For Sa -- Sa mergers at ${\rm z \gta 1}$, the increase of metallicity with decreasing redshift is seen to be overcome in its reddening effect by the dominant blueing due to the successively lower present ages. Towards ${\rm z \gta 2}$ cluster metallicities get very low implying bluer colors despite their relatively high present ages. 
We conclude that {\bf chances to reveal a secondary GC population in a merger remnant are slightly better in ${\rm V-I}$ than in ${\rm B-V}$. Depending on the redshift and type of the merging galaxies, secondary GCs today may either be bluer or redder in ${\rm V-I}$ than the primary population.} 

In ${\rm (V-K)}$, which in SSP models older than $\sim 2$ Gyr is seen to strongly split up with metallicity while remaining almost constant in time to $> 12$ Gyr, effects are even stronger. Essentially all mergers of earlier spiral types at ${\rm z < 2}$ will leave GCs significantly {\bf redder} than the primary population due to their higher metallicity. High-redshift mergers (${\rm z \gta 1.5 - 2}$) of later spiral types will leave GCs significantly {\bf bluer} than the primary population due to their very low metallicity. Hence, {\bf chances to detect a secondary GC population in merger remnants are significantly better in ${\rm V-K}$ than in the optical colors and secondary GCs again may either be bluer or redder than the primary population}.  

While mergers of early and intermediate spiral types Sa and Sb over the redshift range from ${\rm z \gta 0.5}$ through ${\rm z \sim 2}$ produce secondary GCs with ${\rm V-I \lta 1.1}$, none of our models really reaches the average ${\rm \langle V-I \rangle_{red} \sim 1.2}$ color of the red GC subpopulation in E/S0 galaxies (cf. Sect. 6). The same holds true in ${\rm V-K}$, for which our models reach $\sim 3.0$ at maximum, while the red peak GCs in M87, NGC 3115, NGC 4365, and NGC 5846 are reported to show ${\rm \langle V-K \rangle_{red} =3.2,~3.0,~3.2,~and~3.2}$, respectively by Kissler - Patig \etal (2002), Puzia \etal (2002), and Hempel \etal (2002). This slight discrepancy is due to the fact that our models can only give lower limits to the metallicity of secondary GCs in three respects: first, our models do not consider any enrichment during the burst but assume for simplicity that all clusters form with the gaseous metallicity at the onset of the burst. Within burst durations of order $10^8$ yr some degree of additional enrichment may well happen, although the cooling timescales for the gas are difficult to assess. Second, our galaxy model metallicities refer to the average gas phase abundance in an average luminosity spiral of its type. Spirals more luminous than average will have higher gaseous abundances. Third, spirals are known to have metallicity gradients and if secondary GC formation is concentrated to the inner regions and does not involve much gas from beyond 1 effective radius of the disks, the metallicity of our secondary GCs will also be underestimated. 

\subsection{Luminosity Evolution}
Similarly as in Fig. 4 for the colors we present in Fig. 5 the luminosities in V and K of clusters forming in mergers of various spiral types at different redshifts. Normalisation of all models is such that clusters with [Fe/H]$=-1.35$, the mean metallicity of MW GCs, also show ${\rm \langle M_{V_o} \rangle^{MW}_{GCs} = -7.3}$ at an age of 12 Gyr. 

\begin{figure}[!htb]
\centerline{\vbox{
\psfig{figure=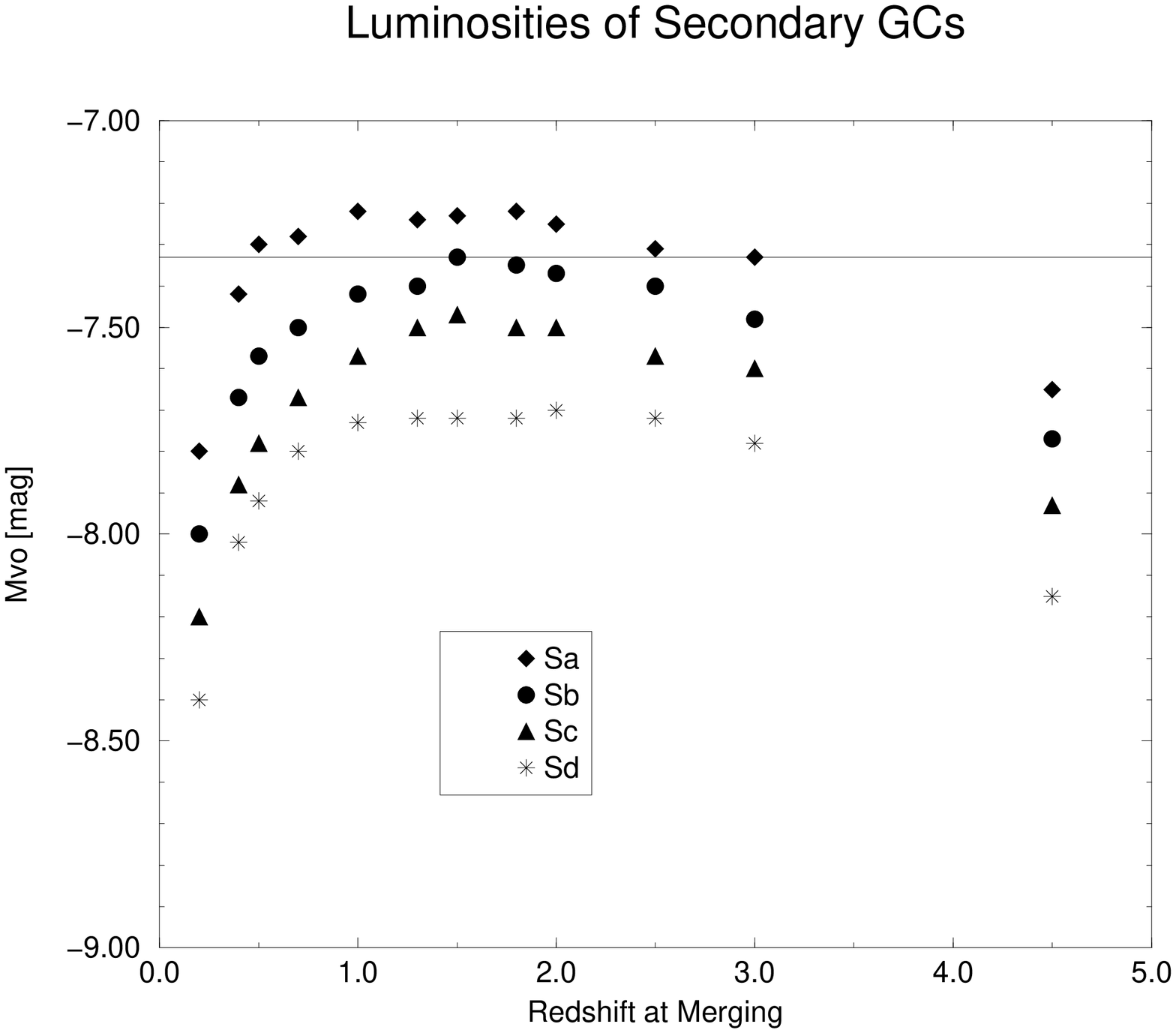,width=6.5cm,angle=270}
\psfig{figure=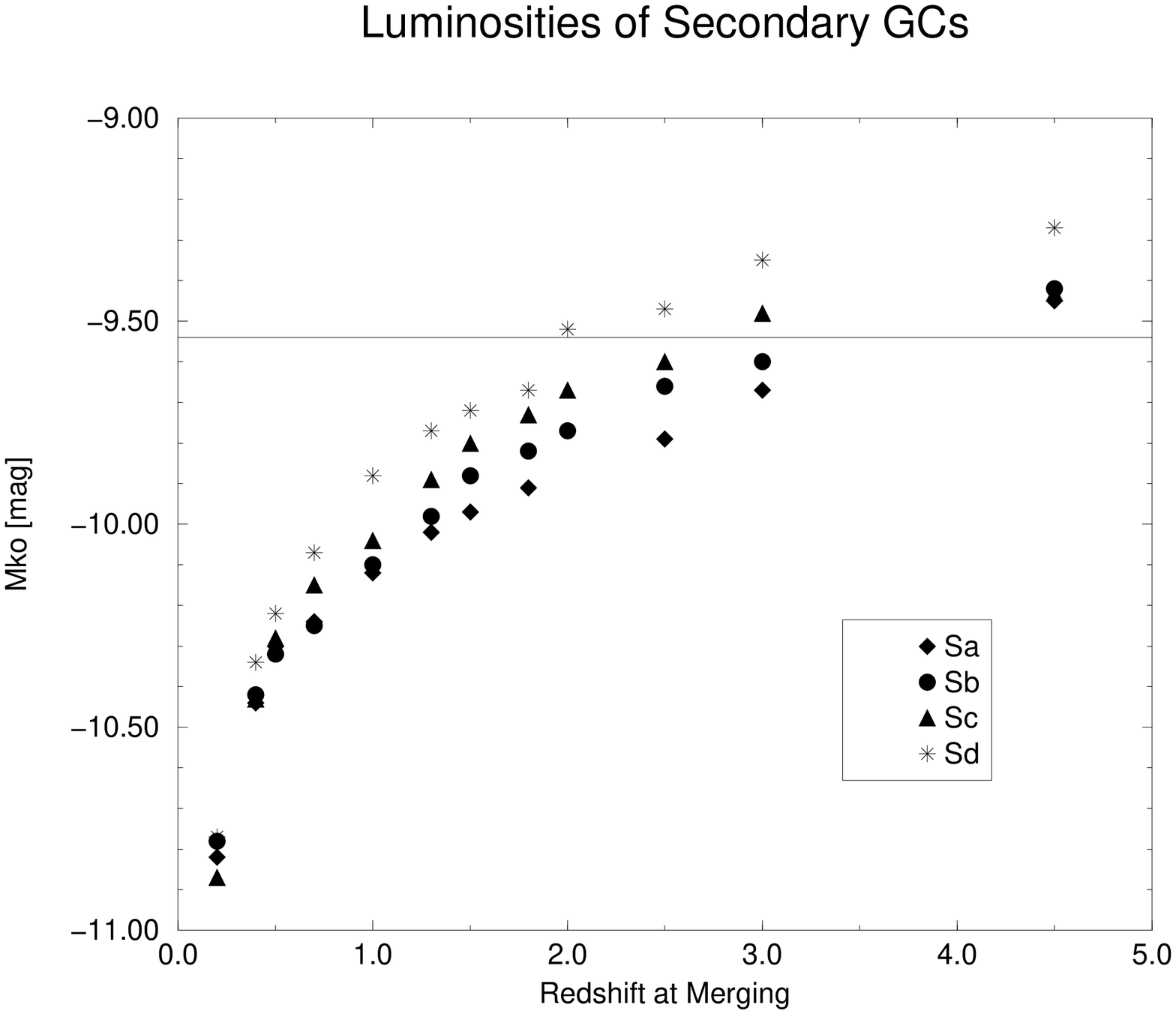,width=6.5cm,angle=270}
}}
\caption{Present day luminosities in V and K of secondary GCs formed in mergers of various galaxy types occuring at different redshifts. Horizontal lines mark mean ${\rm \langle M_{V_o} \rangle}$ and ${\rm \langle M_{K_o} \rangle}$ of MW GCs. For individual GCs in M31, e.g., observational luminosities are uncertain by $\pm 0.13$ mag in ${\rm M_V}$ and by $\pm 0.20$ mag in ${\rm M_K}$ (cf. Barmby \etal 2001), while the intrinsic dispersion in the MW and M31 GC LFs is ${\rm \sigma(M_V) \sim 1.3~mag}$ (Ashman \etal 1995).}
\end{figure}

Fig. 5 shows that for mergers of any type of galaxies there is a relatively broad redshift range (${\rm 1 \lta z \lta 2}$) that produces clusters with comparable V-band luminosity. Mergers occuring earlier or later than this produce clusters that, by today, are brighter due to lower metallicities or younger ages, respectively. Only early-type spiral mergers (Sa - Sb) in the redshift range ${\rm 0.3 \leq z \leq 3}$ produce clusters which by today show luminosities close to ${\rm \langle M_{V_o} \rangle^{MW}_{GCs}}$. In all other cases, clusters are expected to be brighter than MW GCs in the V-band by today. 

In the K-band, the situation is different. Clusters produced in mergers of all kinds of spirals in the redshift interval ${\rm 2 \leq z \lta 3}$ will have present-day K-luminosities comparable to those of MW GCs. Due to the inverse metallicity trend of K-band luminosity compared to V, almost all mergers at ${\rm z \lta 1.5}$ produce clusters that are significantly brighter than the average MW GC by today. Only very high redshift mergers (${\rm z \gta 3}$) -- of late-type galaxies in particular -- are expected to produce GCs fainter than ${\rm \langle M_{K_o} \rangle^{MW}_{GCs}}$. 

The same reasons as given in the previous sections that imply that our secondary cluster metallicities are lower limits lead us to expect that our V- and K-band luminosities are bright and faint limits, respectively. 

\clearpage

\section{Color and Luminosity Evolution of GC Systems in Merger Remnants}
While in the previous Section, we investigated the mean colors and luminosities of secondary GCs forming in mergers, we now want to compare the color distributions and LFs of secondary GC populations formed in mergers of various spiral types at different redshifts to those of the primary GC population inherited from the merging spirals to find out in which cases bimodal distributions might be detected observationally and in which cases a secondary population may hide among the primary one or just broaden their respective distributions. We will see that this is different for different colors, making multiband imaging including NIR a powerful approach to disentangle GC populations.

\begin{figure*}[!htb]
  \centering \epsfig{file=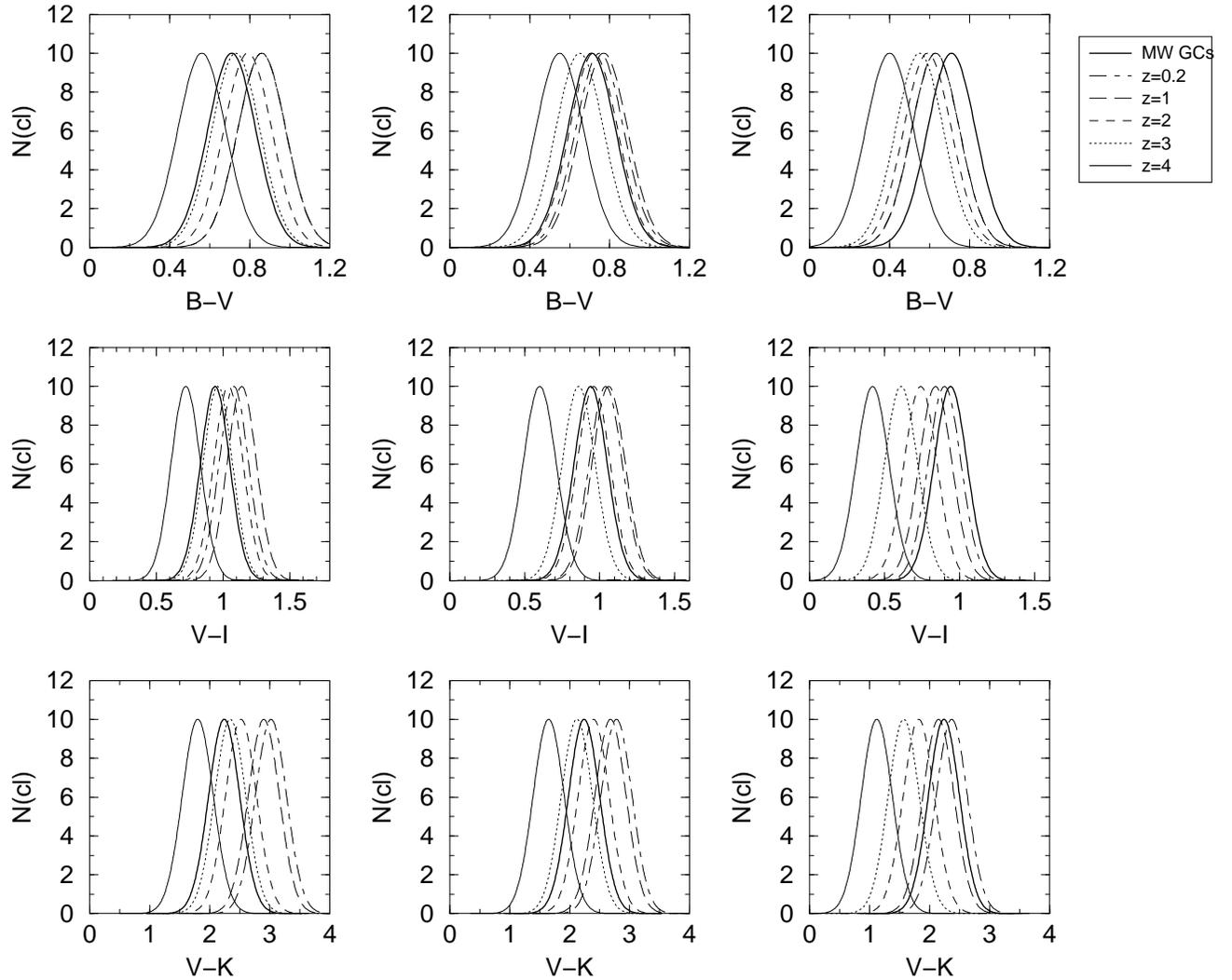,width=14.cm,angle=270}
  \caption{Color Distributions in ${\rm (B-V),~(V-I)~and~(V-K)}$ of secondary GC populations in mergers of various spiral types at different redshifts.  
Sa -- Sa mergers (${\rm 1^{st}}$ column), Sb -- Sb mergers (${\rm 2^{nd}}$ column, Sd -- Sd mergers (${\rm 3^{rd}}$ column). 
The thick solid line shows the corresponding MW halo GC color distributions for 
reference.}
\end{figure*}

\begin{figure*}[!htb]
  \centering \epsfig{file=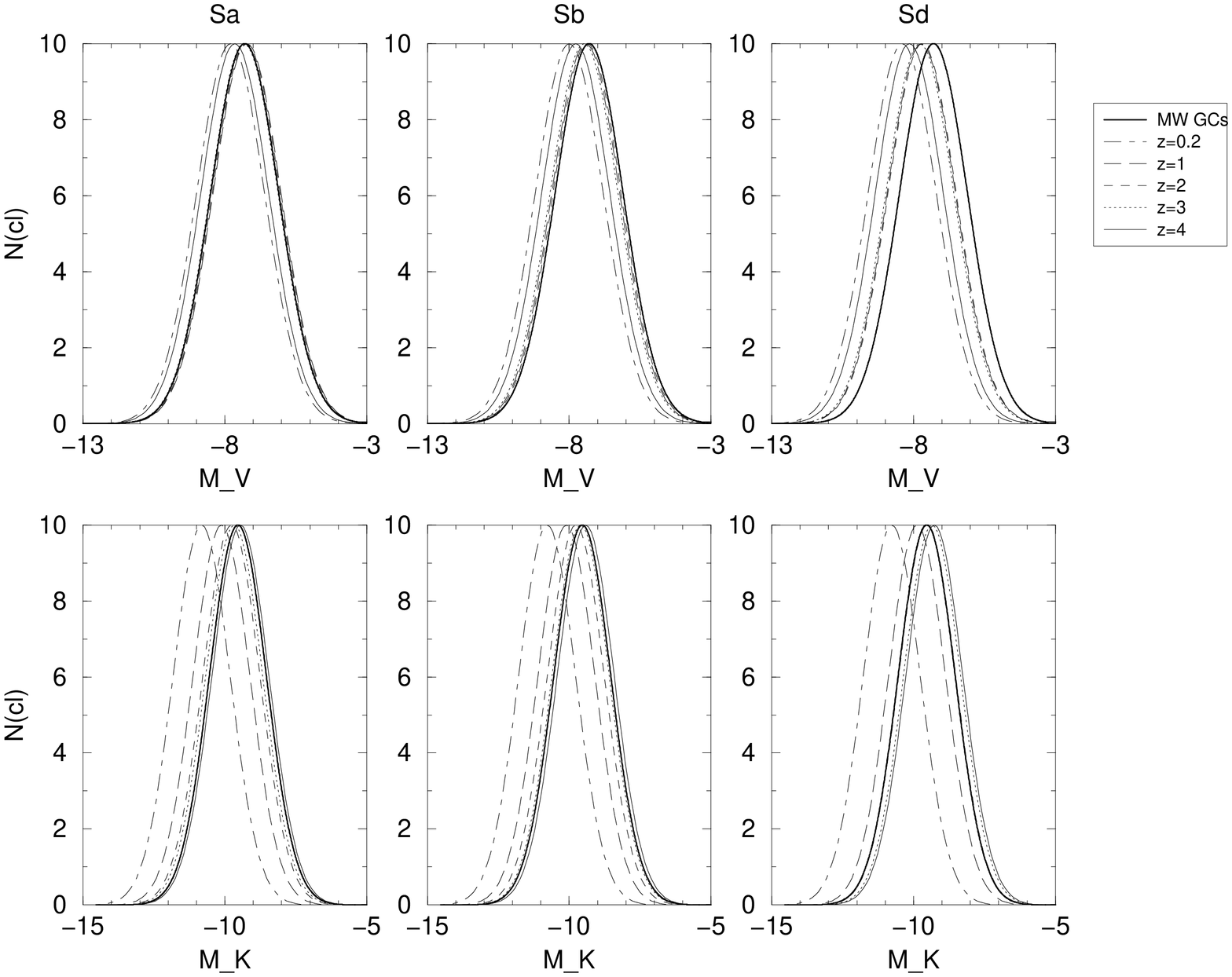,width=13.cm,angle=270}
  \caption{Luminosity Evolution in V (top panels) and K (lower panels) of  secondary GC populations in mergers of various spiral types at different redshifts.  
Sa -- Sa mergers (${\rm 1^{st}}$ column), Sb -- Sb mergers (${\rm 2^{nd}}$ column, Sd -- Sd mergers (${\rm 3^{rd}}$ column). 
The thick solid line shows the corresponding MW GC LFs for 
reference.}
\end{figure*}

As in Sect. 2 we again assume that both GC populations have the same intrinsic widths in their color distributions and LFs and neglect statistical and sampling errors. And again, for the sake of clarity in the presentation and lack of better knowledge we assume that both populations contain comparable numbers of clusters.

\subsection{Color Distributions of GC Systems in Merger Remnants}
In Fig. 6 we present the present-day color distributions in ${\rm (B-V)}$, ${\rm (V-I)}$, and ${\rm (V-K)}$ predicted for secondary GCs that formed in spiral -- spiral mergers at various redshifts. We also plot the observed color distributions of MW halo GCs as a reference for the primary GC population in our mergers. We see that GCs formed in very early mergers (${\rm z \sim 4}$) are always bluer by today than MW GCs in any color, no matter what kind of spiral type was involved in the merger. This is due to the very low metallicity [Fe/H]$\leq -2$ that these GCs have, which is, in fact lower than that of the MW GCs. Born with increasing metallicity towards lower redshift, these secondary GCs then end up redder despite the lower ages they can achieve until ${\rm z=0}$. This effect is, of course, stronger in earlier spiral types because of their stronger enrichment. GCs forming in Sd -- Sd mergers always remain bluer in ${\rm B-V}$ and ${\rm V-I}$ than MW GCs and reach at most the MW GC ${\rm V-K}$ color because of the slow enrichment in those galaxies. 

Fig. 6 shows that the color distributions of secondary GC systems are bluer than those of the MW GCs for all very early mergers at ${\rm z\sim 4}$ and Sd-Sd mergers at ${\rm z \geq 2}$, they are hidden in the primary GC color distributions for Sa-Sa mergers at ${\rm z\sim 3}$, Sb-Sb mergers at ${\rm 3 \geq z \geq 2}$, and Sd-Sd mergers at ${\rm z\sim 0.2}$. Color distributions of secondary GCs finally are redder than those of MW GCs for Sa-Sa mergers at ${\rm z < 2}$ and  Sb-Sb mergers at ${\rm z\leq 1}$, most notably in ${\rm V-K}$. We recall that for reasons detailed in Sect. 4.1. our secondary GC mdel colors are expected to mark the blue limit of the true secondary GC color distributions.

\subsection{Luminosity Functions of GC Systems in Merger Remnants}

In Fig. 7 we show the present-day V- and K-band LFs of secondary GCs formed in mergers of various spiral types at different redshifts. In the V-band secondary GC LFs differ very little from primary ones for Sa and Sb mergers, no matter at which redshift they occur. GCs formed in Sd mergers are brighter than ${\rm \langle M_{V_o} \rangle ^{MW}_{GCs}}$ both for very early mergers  at ${\rm z \sim 4}$ (due to the very low metallicity despite their old ages) and late ones at ${\rm z \sim 0.2}$ (due to their young ages at still rather low metallicity). In K, GCs formed in late mergers, around ${\rm z \sim 0.2}$, are significantly brighter than ${\rm \langle M_{K_o} \rangle ^{MW}_{GCs}}$, no matter what galaxy types were involved. If for reasons discussed above secondary GCs should be more metal-rich than our models assume, we expect their V- and K-band turn-overs to occur at slightly fainter and brighter magnitudes, respectively.

\subsection{Comparing Color Distributions and LFs}
Comparing the LFs for secondary GCs forming in spiral -- spiral mergers in Fig. 7 with those for secondary GCs with any arbitrary combination of age and metallicity in Fig. 2, we see striking differences in both V and K. {\bf Despite its considerable width the age -- metallicity relation of spiral galaxies} (cf. Fig. 3) {\bf acts to exclude most of the cases where the present-day LFs of secondary GC populations strongly differ from the MW GC LFs, both in V and K.} 
This is not true for the color distributions. Comparing color distributions for secondary GCs with arbitrary combinations of age and metallicity (Fig. 1) and those for GCs produced in galaxy mergers (Fig. 6), we see that the manifold of distributions is different, in particular in ${\rm V-K}$, but not drastically reduced like in case of LFs. 

Now let us compare the {\bf color distributions} and LFs for secondary GCs born in galaxy mergers. In contrast to the situation for arbitrary combinations of ages and metallicities (Sect. 2.5) where bimodal LFs were found in cases where the color distributions were either bimodal or even single-peaked, we here see that V-band LFs appear single-peaked in most cases, including those which clearly show bimodal color distributions. Only in case of either extremely early mergers of Sa/Sb galaxies at ${\rm z=3~or~4}$ or of Sd -- Sd mergers -- at all epochs -- a secondary brighter peak might marginally be identified in rich GC systems observed with good photometric accuracy (${\rm 1 \sigma ~errors~ \lta 0.2~mag}$). 

This result is a consequence of the restricted range of age -- metallicity combinations in spiral mergers, i.e. of what we called the cosmic age -- metallicity relation for spiral galaxies. It is very important since it ``saves'' the method of distance determination using GC LFs and, in particular, their ``universal'' turn-over magnitudes ${\rm \langle M_{V_o} \rangle}$ -- at least if the bimodal GC distribution is caused by a spiral -- spiral merger. 

We note that other scenarios for the formation of secondary GCs populations than the spiral -- spiral merger scenario may also leave the LF of the resulting total GC population unaffected, provided the secondary GCs form from material that somehow obeyed an age -- metallicity relation similar to the one we are using here.

In K-band LFs, however, secondary GCs should be detected more easily, at least for rather recent mergers and those that produced high metallicity clusters. 

Again we see that multi-band imaging of large enough GC populations with good photometric accuracy may allow to extract information about cluster formation epochs and metallicities and, hence, shed light on the violent star formation epochs of their parent galaxies. Filter systems particularly sensitive to metallicity (in late evolutionary stages) and age like e.g. the Washington or Stroemgren systems may even be better suited for this purpose. We caution, however, that the conventional calibrations, e.g. of the Washington color C-T1 or of Stroemgren m$_1$ in terms of metallicity [Fe/H], are empirically derived from MW GC data, i.e. valid only for GCs as old as $\sim 12$ Gyr and with metallicities [Fe/H]$ \lta -0.5$. Comparison with theoretical calibrations obtained from SSP evolutionary synthesis models by Schulz \etal (2002) show that 1. for cluster ages $\lta 10$ Gyr the calibrations change significantly and 2. for all ages the relations steepen at [Fe/H]$>-0.5$, for Johnson/HST colors ${\rm (B-V),~(V-I),~(V-K)}$ as well as for Stroemgren ${\rm m_1}$ (but not for Washington C-T1) (cf. Schulz \etal 2002, Fig. 9, 10). Hence, unless a GC system is known to be old in its entirety ($>10$ Gyr), additional NIR data are required to resolve the age-metallicity degeneracy, not only for broad-band colors, but also for Washington or Stroemgren photometry. Our SSP models include these filter systems and allow the results of this paper to be readily extended in this respect. 

\subsection{Future Prospects}
Within the framework of an ESO/ST-ECF ASTROVIRTEL project (PI R. de Grijs) we are currently collecting multi-wavelength LFs and color distributions for GC and young star cluster systems in a large variety of nearby galaxies from {\sl bona fide} old E and S0 galaxies, E/S0s with fine structure or kinematic peculiarities, merger remnants, all through interacting galaxies with ongoing starbursts. This will provide a unique database to explore the universality or environmental dependence of GC LFs, the number of distinct cluster populations as well as their respective metallicities and ages, and should ultimately give information about the violent star formation histories of all these galaxies and, hence, shed light upon galaxy formation scenarios. The work presented here is meant to be the theoretical basis for our analysis of the ASTROVIRTEL data.  

It is also very important to obtain spectra of reasonable numbers of GCs both in systems with bimodal and with rich unimodal color distributions, to reveal if they contain different metallicity subpopulations that -- for specific ages -- may perfectly hide within one inconspicuous peak of the color distribution, as we have shown. This is well feasible now out to more than Virgo cluster distances e.g. with VLT multi-object spectroscopy over large fields and of GCs preselected on HST images.    

We feel that these first and very simplified models provide a useful tool for the interpretation of GC color distributions and LFs. They clearly need to be 
extended to mergers among galaxies of different types, including gas-rich irregular, dwarf, and LSB galaxies. 

\section{First Comparison with Observations}
While the detailed comparison of the models presented here with multi-wavelength data for star cluster systems will be the subject of a series of forthcoming papers (cf. Anders \etal 2003a, b), we present a very first example here for illustration. 

We choose to use some of the V- and I-band data presented by Larsen \etal (2001) for their sample of 17 local early-type galaxies. They find that in nearly all cases two Gaussians allow for a significantly better fit to the ${\rm V-I}$ distributions of the cluster systems than a single one. They have three galaxies in their sample with clear bimodality in the ${\rm (V-I)}$ color distribution, more than 100 GCs in each of the two peaks, and V-band LFs separately determined for GCs from the red and blue peaks: NGC 4472, NGC 4486, and NGC 4649. Their blue peak ${\rm \langle V-I \rangle}$ values span the very small range from 0.943 to 0.954, their red peak ${\rm \langle V-I \rangle}$ values range from 1.196 through 1.207. 
Larsen et al. fit t$_5$-functions to the blue- and red-peak GC populations separately. They chose to use t$_5$-functions instead of Gaussians but show that this does not affect the turn-over magnitudes resulting from the fit. For the 3 galaxies we selected above, they present ${\rm \langle M_V \rangle}$ values for blue- and red-peak clusters obtained from a two-parameter fit (${\rm \langle M_V \rangle,~\sigma \langle M_V \rangle}$) and a one-parameter fit with $\sigma$ fixed at ${\rm \sigma \langle M_V \rangle =1.1}$ mag, the dispersion for MW GCs (Secker 1992). They argue that in view of outlying data points, inaccurate contamination and/or completeness corrections, etc., the one-parameter fit probably gives more accurate turn-over magnitudes. Therefore we use these values for ${\rm \langle M_V \rangle}$. 

In Table 2 we summarize the mean ${\rm (V-K)}$ colors of the blue and red GC populations, ${\rm \langle V-K \rangle_b}$ and ${\rm \langle V-K \rangle_r}$, their respective numbers ${\rm N_b}$ and ${\rm N_r}$, and the differences ${\rm \Delta M_V := \langle M_V \rangle_{red} - \langle M_V \rangle_{blue}}$ in the turn-over magnitudes of the separate LFs of the blue- and red-peak GCs for those three galaxies from Larsen \etal's work.

\begin{table}[htbp]
\begin{center}
\caption{Observed properties of blue- and red-peak GCs in 3 galaxies from Larsen \etal's (2001) sample. ${\rm \Delta M_V}$.}
\begin{tabular}{|c|c|c|c|c|c|}
\hline
 & & & & & \\ 
 NGC & ${\rm \langle V-I \rangle_b}$ & ${\rm \langle V-I \rangle_r}$ & ${\rm N_b}$ & ${\rm N_r}$ & ${\rm \Delta M_V}$ \\
 & & & & & \\
\hline
 & & & & & \\
 4472 & 0.943 & 1.207 & 277 & 255 & 0.5 \\
 4486 & 0.951 & 1.196 & 334 & 375 & 0.22 \\
 4649 & 0.954 & 1.206 & 176 & 169 & 0.26 \\
 & & & & & \\
\hline

\end{tabular}
\end{center}
\end{table}

Assuming an age of 12 Gyr for the blue peak clusters implies metallicities -- as derived from our models -- of [Fe/H]$ =-1.39 ~.~.~.~ -1.43$. If they had the same age of 12 Gyr as the blue clusters, we would derive metallicities for the red clusters in the range $-0.27 ~.~.~.~ -0.33$. Barmby et al.'s (2000) empirical relation for MW GCs leads to [Fe/H]$=-1.36 ~.~.~.~ -1.41$ for the blue and [Fe/H]$=-0.30 ~.~.~.~ -0.39$ for the red GCs, respectively. 

Hence, the ${\rm V-I}$ colors of the blue peaks in all three galaxies are well compatible with an age of 12 Gyr and metallicities very close to the mean metallicity of the M31 halo GCs ${\rm \langle [Fe/H]\rangle = -1.43}$, as determined by Barmby et al. (2000). Raising the ages of the blue peak clusters from 12 to 14 Gyr would decrease their metallicities to $-1.47 \dots -1.53$, i.e. close to those of the Milky Way halo GCs. 

Whether it is reasonable to assume identical ages for two GC populations clearly distinct in ${\rm V-I}$ is certainly a matter of debate. We therefore explore the range of age -- metallicity pairs compatible with a red ${\rm V-I}$ color peak around ${\rm \langle V-I \rangle_r \sim 1.2}$. Pushing to the extreme, the red GC population could have a metallicity as high as [Fe/H]$=+0.4$ and an age as young as $\sim 2.5$ Gyr. At solar metallicity, it would be $\sim 4$ Gyr young and at [Fe/H]$=-0.4$ it may have ages in the range 7 -- 13 Gyr because the ${\rm V-I}$ evolution of the SSP is very flat in this age range. 

{\bf Our models, however, predict both significantly different ${\rm \langle V-K \rangle}$ colors and turnover magnitudes ${\rm \langle M_V \rangle}$ for the red GCs at the different age -- metallicity combinations allowed by their ${\rm \langle V-I \rangle_r}$.} 
In Table 3 we present the predictions of our models for the turn-over differences ${\rm (\Delta M_V)_{pred}}$ and for the mean ${\rm \langle V-K \rangle_{pred}}$ colors of the blue and red GC subpopulations for the different combinations of metallicity and age compatible with Larsen \etal's ${\rm \langle V-I \rangle_r}$. 

In the case of equal age blue- and red-peak clusters, our models yield a difference of 0.4 mag in their turn-over magnitudes. In all three galaxies, NGC 4472, NGC 4486, NGC 4649, the red-peak GCs are observed to be {\bf fainter} than the blue-peak ones by ${\rm \Delta M_V =~0.5,~0.2}$, and 0.3 mag, respectively (cf. Table 2). 
This may be used to rule out the metal-rich and young red cluster solutions ${\rm (+0.4,~0.25~Gyr)}$ and ${\rm (0,~4~Gyr)}$ since their model ${\rm \langle M_V \rangle}$ are predicted to be {\bf brighter} by 0.3 and 0.1 mag, respectively (cf. Table 3). Most of the red cluster solutions ${\rm (-0.4,~7 \dots 13~Gyr)}$ implying that red clusters should be fainter than blue ones by 0.1 (7 Gyr) to 0.6 mag (13 Gyr) are compatible with the observed turn-over differences. This clearly shows that {\bf the availability of distinct LFs for the blue and red GC subpopulations significantly reduces the number of acceptable metallicity -- age combinations}, as already pointed out by Puzia \etal (1999), provided that both subpopulations have the same characteristic cluster masses. 

\begin{table}[htbp]
\begin{center}
\caption{Predicted properties for blue- and red-peak GCs from our models for different possible combinations of metallicity and age ([Fe/H], age [Gyr]) for the red-peak GCs.}
\begin{tabular}{|cc|c|cc|}
\hline
 & & & & \\ 
 \multicolumn{2}{|c|}{([Fe/H], age [Gyr])} & ${\rm (\Delta M_V)_{pred}}$ & \multicolumn{2}{|c|}{${\rm \langle V-K \rangle_{pred}}$} \\
 & & & & \\
\hline
 & & & & \\
 Blue & Red & & Blue & Red \\
 & & & & \\
 $(-1.4,~12)$ & $(-0.3,~12)$ & $0.4$ & 2.4 & 3.2 \\
 $(-1.4,~12)$ & $(+0.4,~2.5)$ & $-0.3$ & 2.4 & 3.6 \\
 $(-1.4,~12)$ & $(~~~0,~~4)$ & $-0.1$ & 2.4 & 3.3 \\
 $(-1.4,~12)$ & $(-0.4,~~7)$ & $0.1$ & 2.4 & 3.0 \\
 $(-1.4,~12)$ & $(-0.4,~13)$ & $0.6$ & 2.4 & 3.1 \\
 & & & & \\
\hline

\end{tabular}
\end{center}
\end{table}

While our models predict ${\rm \langle V-K \rangle_b \sim 2.4}$ for the blue-peak GCs at ages 12 . . . 14 Gyr and metallicities [Fe/H]$=-1.4 .\:.\:.-1.5$, their predictions for ${\rm \langle V-K \rangle_r}$ for the red-peak GCs diverge significantly among the different metallicity -- age pairs that are compatible with the observed ${\rm \langle V-I \rangle_r}$. 

If the red-peak GCs are 12 Gyr old as the blue ones and have [Fe/H]$=-0.3$, our models predict they should have ${\rm \langle V-K \rangle = 3.2}$. If they are very metal-rich and young ([Fe/H]$=+0.4$, age $=2.5$ Gyr), they should show ${\rm \langle V-K \rangle = 3.6}$. At solar metallicity and an age of 4 Gyr they should have ${\rm \langle V-K \rangle = 3.3}$ and at [Fe/H]$=-0.4$ and ages of 7 -- 12 Gyr they are expected to show  ${\rm \langle V-K \rangle = 3.0}$. 
Hence, {\bf K-band observations should allow to significantly further reduce the range of possible age -- metallicity combinations}, no matter if or if not the red- and blue-peak GCs have the same characteristic masses. 

\section{Conclusions}
In the local Universe, the formation of populous systems of compact bright star clusters is a characteristic feature of vigorous starbursts generally accompanying gas-rich galaxy mergers. At least some fraction of these star clusters probably survives for many Gyr and forms a secondary GC population in addition to the primary one(s) inherited from the merged galaxies. The metallicity of a secondary GC system is determined by the enrichment of the gas in the interacting or starburst galaxy at the time of the burst. Hence, we expect GC systems to record valuable information about the violent star formation episodes in the history of their galaxies. 

Different galaxy formation scenarios imply different predictions for the age and metallicity distributions of GCs. In the monolithic early collapse scenario, we expect essentially one population of GCs, all of similar age and metallicity. If one or a few major gas-rich mergers have built up a galaxy we expect at least two different GC populations. Purely stellar mergers would only combine the primary GC systems of their progenitor galaxies. If a galaxy is formed by a prolonged accretion process from a number of subgalactic fragments, multiple GC populations may result and not necessarily a bimodal GC color distribution. 

Color distributions of the majority of GC systems around luminous early-type galaxies are found to be bimodal, their blue peaks seem to be fairly universal and very similar to that of the Milky Way halo GCs. 
Luminosity functions of GC systems around early-type galaxies, however, are conventionally assumed to feature one fairly uniform turn-over that can be used for distance measurements and determinations of the Hubble constant. The age -- metallicity degeneracy of broad band colors precludes a direct conversion from color to metallicity. GC subpopulations of different ages and metallicities may be hidden within one color peak, and it is {\sl a priori} not clear if GCs in a red peak are older and as metal-poor as those in the blue peak or younger and more metal-rich than those. 

In a first step, results from evolutionary synthesis models for cluster populations of various metallicitites -- assuming an intrinsic width of the color distribution constant in time at the value observed for the Milky Way and M31 -- show how their color peak moves from blue to red, depending on metallicity, as they age, in comparison with the blue peak of the Milky Way halo GC population. We find that color differences to the MW halo GC reference peak are very different in different colors and that the NIR is very useful in splitting up two GC populations of different ages and metallicities that in the optical are hidden within one color peak. Hence, accurate multi-wavelength photometry, including NIR, of GC systems will allow to largely disentangle different age and metallicity GC subpopulations. 

We also calculate the evolution of GC LFs in various passbands for GC systems of various metallicities assuming that the width of a luminosity function is constant in time and given by a universal Gaussian GC mass function like the one observed for the MW GCs. Again, we show how the peaks of the LFs move from brighter to fainter magnitudes as a GC population ages, in a way that depends on the metallicity of the GC system and on the wavelength band observed. In comparison with the MW GC LF we identify combinations of age and metallicity for which one inconspicuous peak, a broadened one, or two peaks might be detected. 

Comparing color distributions and LFs of secondary GCs with arbitrary combinations of metallicities and ages with those of the MW halo GCs we find that all possible combinations may occur: inconspicuous single-peak (optical) colour distributions {\bf and} LFs, single-peak color distributions with bimodal LFs, bimodal color distributions with single-peak LFs, as well as both -- color distributions and LFs -- bimodal. Neither bimodal nor particularly broad GC LFs have been reported so far for elliptical or S0 galaxies. They would be a serious problem for distance measurements and the determination of ${\rm H_o}$. 

In a second step we use a very basic age -- metallicity relation for the ISM in normal spiral galaxies to predict metallicities of secondary GCs forming in spiral -- spiral mergers over cosmological times. This relation describes the successive chemical enrichment of the ISM in simple spiral galaxy models. It is broad, extending from low-metallicity late-type spirals to high-metallicity early-type ones at any given age, and has been shown earlier to provide agreement with the redshift evolution of the neutral gas abundances in Damped Ly$\alpha$ absorbing galaxies as well as to naturally link those to the characteristic HII region abundances in various local spiral galaxy types. On the basis of this broad age -- metallicity relation for the ISM in spirals we calculate the color and luminosity evolution of secondary GC populations assumed to form in the strong starbursts accompanying spiral -- spiral mergers at various redshifts. 

Again, we study for which of all those mergers a second GC population would be detectable  in terms of broad band colors against those of the primary population from the progenitor galaxies and we check their LFs. We find that, due to their relatively low metallicities, GCs formed in late-tpe spiral mergers will result in a bluer peak in color distributions if they occur at ${\rm z \sim 2-3}$, but hide within the color distribution of the primary GCs for more recent mergers. Only for mergers of early spiral types (Sa - Sb) at ${\rm z \lta 1 - 2}$, the secondary GCs will populate a distinctly redder peak. 

Intriguingly, the assumption of an age -- metallicity relation for the gas in spirals reduces the manifold of possible combinations in the shapes of color distributions and LFs. It eliminates the bimodal LFs and only leaves single-peak LFs, also in cases where the color distributions clearly are bimodal. This is the situation observed e.g. in NGC 4472 and M87 ($=$ NGC 4486) and thus ``saves'' the method of distance measurement using GC LFs. We note that this is also true for secondary GC formation in other scenarios than that of spiral -- spiral mergers discussed here, provided that the secondary GCs follow an age -- metallicity relation similar to the one adopted here.

For illustration we show as a very first example the application of our model results to the V--I color distributions and V-band LFs of NGC 4472, NGC 4486, and NGC 4649 presented by Larsen et al. (2001). While the availability of only one color leaves a wide range of possible age--metallicity combinations for the GCs in the red peak of these galaxies, this wide range is already narrowed down by the observed differences in the V-band LFs of the blue- and red-peak GCs. V--K colors predicted by our models diverge significantly among the remaining possible age--metallicity combinations. Additional K-band observations are thus expected to finally resolve the age--metallicity degeneracy and allow ages and metallicities of the secondary GC populations to be separately determined, hence, shedding light on the formation epochs of their parent galaxies and the pre-enrichment histories of the building blocks. 

The results presented here are currently used in the interpretation of multi-color GC LFs and color distributions provided by an ESO/ST-ECF ASTROVIRTEL project (PI R. de Grijs).


\acknowledgement
I thank my anonymous referee for a very detailed report that helped improve the presentation of my results. Excellent support given by ASTROVIRTEL, a Project 
funded by the European Commission under FP5 Contract No. HPRI-CT-1999-00081, is gratefully acknowledged. \\


\begin{thebibliography}{}
\bibitem{} Anders, P. \& Fritze - v. Alvensleben, U., 2003, \aap\ 401, 1063
\bibitem{} Anders, P., Bissantz, N., Fritze - v. Alvensleben, U., de Grijs, R., 2003a, \mnras\ {\sl in press}
\bibitem{} Anders, P., de Grijs, R., Fritze - v. Alvensleben, U., Bissantz, N., 2003b, \mnras\ {\sl in press}
\bibitem{} Ashman, K. M. \& Zepf, S. E., 1998, {\sl Globular Cluster Systems}, Cambridge Univ. Press
\bibitem{} Ashman, K. M., Bird, C. M., Zepf, S. E., 1994, \aj\ 108, 2348
\bibitem{} Ashman, K. M., Conti, A., Zepf, S. E., 1995, \aj\ 110, 1164
\bibitem{} Barmby, P., Huchra, J. P., Brodie, J. P., Forbes, D. A., Schroeder,  L. L., Grillmair, C. J., 2000, \aj\ 119, 727
\bibitem{} Barmby, P., Huchra, J. P., Brodie, J. P., 2001, \aj\ 121, 1482
\bibitem{} Brodie, J. P., Larsen, S. S., Kissler -- Patig, M., 2000, \apj\ 543, L19
\bibitem{} Carlson, M. N. \& Holtzman, J. A., 2001, \pasp\ 113, 1522
\bibitem{} C\^ot\'e, P., 1999, \aj\ 118, 406
\bibitem{} C\^ot\'e, P., Marzke, R. O., West, M. J., Minniti, D., 2000, \apj\ 533, 869
\bibitem{} Couture, J., Harris, W. E., Allwright, J. W. B., 1990, \apj\S 73, 671
\bibitem{} Elson, R. A. W., 1997, \mnras\ 286, 771
\bibitem{} Forbes, D. A., Brodie, J. P., Grillmair, C. J., 1997, \aj\ 113, 1652
\bibitem{} Fritze -- v. Alvensleben, U., 1998, \aap\ 336, 83
\bibitem{} Fritze -- v. Alvensleben, U., 1999a, \aap\ 342, L25
\bibitem{} Fritze -- v. Alvensleben, U., 1999b, in {\sl Spectrophotometric Dating of Stars and Galaxies}, eds. I. Hubeny, S. R. Heap, R. H. Cornett, ASP Conf. Ser. 192, 273
\bibitem{} Fritze -- v. Alvensleben, U., 2000, in {\sl Massive Stellar Clusters}, eds. A. Lan\c{c}on \& C. M. Boily, ASP Conf. Ser. 211, 3
\bibitem{} Fritze -- v. Alvensleben, U., 2001, in {\sl The Evolution of Galaxies: I. Observational Clues}, eds. J. M. V\'ilchez, G. Stasi\'nska, E. P\'erez, Kluwer, Dordrecht, p. 305
\bibitem{} Fritze -- v. Alvensleben, U. \& Burkert, A., 1995, \aap\ 300, 58
\bibitem{} Fritze -- v. Alvensleben, U. \& Gerhard, O. E., 1994a, \aap\ 285, 751
\bibitem{} Fritze -- v. Alvensleben, U. \& Gerhard, O. E., 1994b, \aap\ 285, 775
\bibitem{} Gebhardt, K. \& Kissler -- Patig, M. 1999, \aj\ 118, 1526 
\bibitem{} Geisler, D. \& Forte, J. C., 1990, \apj\ 350, L5
\bibitem{} Geisler, D., Lee, M. G., Kim, E., 1996, \aj\ 111, 1529
\bibitem{} Harris, W. E., 1991, \araa\ 29, 543
\bibitem{} Harris, G. L. H., Harris, W. E., Poole, G. B., 2000, \aj\ 117, 855
\bibitem{} Hempel, M., Hilker, M., Kissler -- Patig, M., Puzia, T. H., Minniti, D., Goudfrooij, P., 2002, \aap\ 405, 487
\bibitem{} Kavelaars, J. J., Harris, W. E., Hanes, D. A., Hesser, J. E., 
Pritchet, C. J., 2000, \apj\ 533, 125
\bibitem{} Kissler -- Patig, M., Grillmair, C. J., Meylan, G., Brodie, J. P., Minniti, D., Goudfrooij, P., 1999, \aj\ 117, 1206
\bibitem{} Kissler -- Patig, M., Brodie, J. P., Minniti, D., 2002, \aap\ 391, 441
\bibitem{} Kundu, A. \& Whitmore, B. C., 1998, \aj\ 116, 2841
\bibitem{} Kundu, A. \& Whitmore, B. C., 2001a, \aj\ 121, 2950
\bibitem{} Kundu, A. \& Whitmore, B. C., 2001b, \aj\ 122, 1251
\bibitem{} Kurth, O. M., Fritze -- v. Alvensleben, U., Fricke, K. J., 1999, \aap\ 138, 19
\bibitem{} Larsen, S. S., Brodie, J. P., Huchra, J. P., Forbes, D. A., Grillmair, C. J., 2001, \aj\ 121, 2974
\bibitem{} Le F\`evre, O., Abraham, R., Lilly, S. J., Ellis, R. S., \etal 2000, \mnras\ 311, 565
\bibitem{} Lindner, U., Fritze -- v. Alvensleben, U. \& Fricke, K. J., 1999, \aap\ 341, 709
\bibitem{} Maraston, C., Kissler -- Patig, M., Brodie, J. P., Barmby, P., Huchra, J. P., 2001, \aap\ 370, 176
\bibitem{} Meurer, G. R., 1995, \nat\ 375, 742
\bibitem{} Meurer, G. R., Heckman, T. M., Leitherer, C., Kinney, A., Robert, C.,
 Garnett, D. R., 1995, \aj\ 110, 2665
\bibitem{} M\"oller, C. S., Fritze -- v. Alvensleben, U. \& Fricke, K. J., 1999, IAU Symp. 183, 158
\bibitem{} Ostrov, P. G., Forte, J. C., Geisler, D., 1998, \aj\ 116, 2854
\bibitem{} Puzia, T. H., Kissler -- Patig, M., Brodie, J. P., Huchra, J. P., 1999, \aj\ 118, 2734
\bibitem{} Puzia, T. H., Zepf, S. E., Kissler -- Patig, M., Hilker, M., Minniti, D., Goudfrooij, P., 2002, \aap\ 391, 453
\bibitem{} Schulz, J., Fritze -- v. Alvensleben, U. \& Fricke, K. J., 2002, \aap\ 392, 1
\bibitem{} Schweizer, F., 2002, IAU Symp. 207, 630
\bibitem{} Schweizer, F. \& Seitzer, P., 1993 \apj\ 417, L29
\bibitem{} Schweizer, F. \& Seitzer, P., 1998, \aj\ 116, 2206
\bibitem{} Secker, J., 1992, \aj\ 104, 1472
\bibitem{} Vesperini, E., 1997, \mnras\ 287, 915
\bibitem{} Vesperini, E., 1998, \mnras\ 299, 1019
\bibitem{} Vesperini, E., 2000, \mnras\ 318, 841
\bibitem{} Vesperini, E., 2001, \mnras\ 322, 247
\bibitem{} Whitmore, B. C., Sparks, W. B., Lucas, R. A., Macchetto, F. D., Biretta, J. A., 1995, \apj\ 454, L73
\bibitem{} Worthey, G., 1994, \apj\S 95, 107
\bibitem{} Zepf, S. E., 2002, IAU Symp. 207, 294
\bibitem{} Zhang, Q. \& Fall, S. M., 1999, \apj\ 527, L81


\end{thebibliography}
\end{document}